\documentclass[pre,twocolumn,datetime]{revtex4-1}  
\usepackage{amsfonts}
\usepackage{amsmath}
\usepackage{amssymb}
\usepackage{version}
\usepackage{graphicx}%
\includeversion{New_connection}
\excludeversion{Old_connection}

\begin{document}
\noindent
Journal of Statistical Physics {\bf 190}, 96 (2023)\\ http://doi.org/10.1007/s10955-023-03104-8
\title{On the Application of Non-Gaussian Noise in Stochastic Langevin Simulations}

\author{Niels Gr{\o}nbech-Jensen}
\email{ngjensen@math.ucdavis.edu}
\affiliation{Department of Mechanical \& Aerospace Engineering and Department of Mathematics\\ University of California, Davis, CA 95616, U.S.A.}

\begin{abstract}



\noindent
In light of recent advances in time-step independent stochastic integrators for Langevin equations, we revisit the considerations for using non-Gaussian distributions for the thermal noise term in discrete-time thermostats. We find that the desirable time-step invariance of the modern methods is rooted in the Gaussian noise, and that deviations from this distribution will distort the Boltzmann statistics arising from the fluctuation-dissipation balance of the integrators. We use the GJ stochastic Verlet methods as the focus of our investigation since these methods are the ones that contain the most accurate thermodynamic measures of existing methods. Within this set of methods we find that any distribution of applied noise, which satisfies the two first moments given by the fluctuation-dissipation theorem, will result in correct, time-step independent results that are generated by the first two moments of the system coordinates. However, if non-Gaussian noise is applied, undesired deviations in higher moments of the system coordinates will appear to the detriment of several important thermodynamic measures that depend especially on the fourth moments. The deviations, induced by non-Gaussian noise, become significant with the one-time-step velocity attenuation, thereby inhibiting the benefits of the methods. Thus, we conclude that the application of Gaussian noise is necessary for reliable thermodynamic results when using modern stochastic thermostats with large time steps.
\end{abstract}
\maketitle
\section{Introduction}
\label{sec:intro}
Over the past several decades, discrete-time Langevin and Brownian simulations in computational statistical mechanics have been a centerpiece in many specific simulation applications to molecular dynamics, e.g., materials science, soft matter, and biomolecular modeling \cite{AllenTildesley,Frenkel,Rapaport,Hoover_book,Leach}. Inherent to simulations is the fundamental complication of time discretization, which on one hand is necessary for advancing time, but on the other hand inevitably corrupts the desired, continuous-time system features of interest. An integral part of a simulation task is therefore to explore and optimize the balance between the two conflicting core objectives; namely simulation efficiency by increasing the time step, and simulation accuracy by decreasing it.

Since discrete-time algorithms for time evolution produce only approximate solutions to a continuous-time solution, and since the errors must vanish in the limit of infinitesimal time step, most considerations for a useful simulation environment have been considered in the limit of small time steps, where the simulated system variables change only little in each step. In that limit, the discrete-time system behaves much like the continuous-time system, and algorithm development has therefore mostly been conducted by formulating objectives and measures in continuous-time, whereafter time has been discretized and executed with relatively small steps in order to retain accuracy.

For stochastic differential equations, notably Langevin equations, an added complication is that it is not a specific trajectory that is of interest, but instead the statistics of trajectories that determines if the desired continuous-time behavior has been adequately represented. The traditional approach to these problems has therefore been to acquire accurate statistics from accurate trajectories obtained from small time steps, where continuous-time behavior is well approximated. However, with the recent advent of discrete-time stochastic algorithms that can produce correct statistics from inaccurate, large time-step trajectories \cite{GJF1,2GJ,GJ}, the details of, e.g., stochastic noise correlation and distribution now become essential for obtaining the advantages of the statistical accuracy for all meaningful time steps. The significance of these details arises from the discretization of time, as outlined in the following.

We are here concerned with the Langevin equation \cite{Langevin,Langevin_Eq}
\begin{eqnarray}
m\dot{v}+\alpha\dot{r} & = & f+\beta \; , \label{eq:Langevin}
\end{eqnarray}
where $m$ is the mass of an object with spatial (configurational) coordinate $r$ and velocity (kinetic) coordinate $v=\dot{r}$. The object is subjected to a force $f=-\nabla E_p(r)$, with $E_p(r)$ being a potential energy surface for $r$, and linear friction that is represented by the non-negative constant $\alpha$. The fluctuation-dissipation relationship specifies that the thermal fluctuations $\beta$ can be represented by a distribution for which the following first two moments are given \cite{Parisi}:
\begin{subequations}
\begin{eqnarray}
\langle\beta(t)\rangle & = & 0 \\
\langle\beta(t)\beta(t^\prime)\rangle & = & 2\alpha\, k_BT\, \delta(t-t^\prime) \; , 
\end{eqnarray}
\label{eq:FD}\noindent
\end{subequations}
where $\delta(t)$ is Dirac's delta function signifying that the fluctuations are temporally uncorrelated, $k_B$ is Boltzmann's constant, and $T$ is the thermodynamic temperature of the heat-bath, to which the system is coupled through the friction constant $\alpha$.
A feature in statistical mechanics is that the fluctuation-dissipation theorem for the noise distribution $\beta(t)$ only specifies the two moments given in Eq.~(\ref{eq:FD}) \cite{Parisi,Langevin_Eq,comment_symmetry}. Thus, there is a great deal of freedom in choosing the distribution of the noise in the Langevin equation such that correct thermodynamics is represented. The reason that the physical (statistical) results do not depend on this freedom is that continuous-time demands a new, uncorrelated noise value $\beta(t)$ be contributed and accumulated at every instant, resulting in a Gaussian outcome at any non-zero time-scale due to the central limit theorem (see, e.g., Ref.~\cite{central_limit}). As a result, the sampled (one-dimensional) Maxwell-Boltzmann velocity distribution $\rho_{k,e}(v)$ and the configurational Boltzmann distribution $\rho_{c,e}(r)$ always become \cite{Langevin_Eq}
\begin{subequations}
\begin{eqnarray}
\rho_{k,e}(v) & = & \sqrt{\frac{m}{2\pi k_BT}}\exp\left(-\frac{\frac{1}{2}mv^2}{k_BT}\right) \label{eq:MBoltzmann_exact}\\
\rho_{c,e}(r) & = & {\cal C}_c\exp\left(-\frac{E_p(r)}{k_BT}\right) \, , \label{eq:BBoltz}
\end{eqnarray}\label{eq:Boltz}\noindent
with the constant ${\cal C}_c$ determined by the normalization of probability
\begin{eqnarray}
\underset{\rm all \; space}{\int} \!\!\!\rho_{c,e}(r)\,dr & = & 1\, , 
\end{eqnarray}
\end{subequations}
regardless of which temporally uncorrelated distribution $\beta(t)$ is at play in Eq.~(\ref{eq:Langevin}), and for as long as Eq.~(\ref{eq:FD}) is satisfied. Similarly, it is given that diffusive transport for a flat surface is represented by a diffusion constant \cite{AllenTildesley}
\begin{eqnarray}
D_E & = & \frac{1}{d}\lim_{t\rightarrow\infty}\frac{\langle[r(t)-r(0)]^2\rangle}{2t}   \label{eq:Einstein_D} \\
& = & \frac{k_BT}{\alpha} \label{eq:_D_E_cont} \, ,
\end{eqnarray}
where $d$ is the dimension of the configurational space.

Given that the central limit theorem ensures a Gaussian noise distribution for any appreciable integration of the instantaneous variable $\beta(t)$, it is natural to choose $\beta(t)$ to be the Gaussian that conforms to Eq.~(\ref{eq:FD}). However, in discrete-time simulations it is not uncommon to directly employ a uniform distribution that satisfies Eq.~(\ref{eq:FD}). The rationale is rooted in the computational efficiency of uniformly distributed pseudo-random generators \cite{random}, compared with the additional computational load required for transforming those variables into Gaussian distributed variables before use. A thorough investigation of using non-Gaussian valuables in discrete time was presented in the 1988 publication Ref.~\cite{Greiner_1}, where it was concluded that, given the inherent inaccuracies of the stochastic discrete-time integrators of the time, the scaling in statistical moments upon the applied time step did not change adversely when using non-Gaussian noise distributions, specifically the uniform distribution. Building upon this work, Ref.~\cite{Dunweg_1} extended the investigation in 1991 to a broader class of noise distributions, which would yield indistinguishable statistical results compared to the use of Gaussian noise for vanishing time steps, where the underlying systemic time-step errors of the then contemporary methods would vanish as well. The studies highlighted the benefits of computational efficiencies in using simple, non-Gaussian noise given that the statistical differences were deemed to be insignificant for small time steps, which in any case were necessary for the accuracy of the stochastic integrators, even using Gaussian noise. As pointed out in a recent article \cite{Finkelstein_1} that compares the quality of contemporary stochastic integrators, this practice has been widespread when using traditional algorithms, such as the Br{\"u}nger-Brooks-Karplus \cite{BBK} (also analyzed in Ref.~\cite{Pastor_88}) or the Schneider-Stoll \cite{SS} method, for as long as small time steps are applied (for a more expansive list of traditional and recent Langevin integrators, please see Ref.~\cite{GJ}). For example, these two methods are used with non-Gaussian noise in, e.g., the wide-distribution molecular dynamics simulation suites, LAMMPS \cite{Plimpton,Plimpton2,LAMMPS_uniform} and ESPResSo \cite{Espresso_4.2.0}.

It is the aim of this work to revisit the discussion of applying non-Gaussian noise in light of the most modern stochastic integrators, which have been designed to produce accurate, time-step independent statistics and transport through the use of Gaussian noise. We demonstrate that even if the low order moments of the configurational and kinetic distributions are seemingly unaffected by the details of the noise distribution, there may still be significant differences in the sampled phase space that affect the statistical sampling of the system. We base the analysis exclusively on the recently formulated GJ set of methods \cite{GJ}, since these methods have the most optimized statistical properties that are invariant to the time step for linear systems when using Gaussian noise. This also allows us to investigate the large time-step implications of using non-Gaussian noise without being subject to other inherent time-step errors of the integrators. The presentation is organized as follows. Section~\ref{sec:GJ_review} reviews the basic features and structure of the GJ methods formulated in both standard Verlet and split operation forms, Sec.~\ref{sec:moments} analyzes analytically the first four statistical moments of the configurational and kinetic coordinates for linear systems, Sec.~\ref{sec:SimDisc} presents companion simulations to the analysis conducted in Sec.~\ref{sec:moments}, and Sec.~\ref{sec:discussion} summarizes the discussion with specific conclusions.

\section{Review of the GJ methods}
\label{sec:GJ_review}
We adopt the complete set of GJ methods \cite{GJ} for solving Eq.~(\ref{eq:Langevin}) in the form
\begin{subequations}
\begin{eqnarray}
u^{n+\frac{1}{2}} & = & \sqrt{c_1}\,v^{n}+\frac{\sqrt{c_3}\,\Delta{t}}{2m}\,f^n+\frac{\sqrt{c_3}}{2m}\,\beta^{n+1} \label{eq:GJ_half}\\
r^{n+1} & = & r^n + \sqrt{c_3}\,\Delta{t}\,u^{n+\frac{1}{2}} \label{eq:GJ_pos}\\
v^{n+1} & = & \frac{c_2}{\sqrt{c_1}}\,u^{n+\frac{1}{2}}+\sqrt{\frac{c_3}{c_1}}\,\frac{\Delta{t}}{2m}\,f^{n+1}+\sqrt{\frac{c_3}{c_1}}\,\frac{1}{2m}\,\beta^{n+1} \, , \nonumber \\ \label{eq:GJ_full}
\end{eqnarray} \label{eq:GJ_combo}\noindent
\end{subequations}
where a superscript $n$ on the variables $r$, $v$, $u$, and $f$ pertains to the discrete time $t_n=t_0+n\,\Delta{t}$ at which the numerical approximation is given, with $\Delta{t}$ being the discrete time step. Thus, $f^n=f(t_n,r^n)$. The two represented velocities are, respectively, the half-step velocity $u^{n+\frac{1}{2}}$ at time $t_{n+\frac{1}{2}}$, and the on-site velocity $v^{n}$ at $t_n$, given by
\begin{subequations}
\begin{eqnarray}
u^{n+\frac{1}{2}} & = & \frac{r^{n+1}-r^n}{\sqrt{c_3}\,\Delta{t}} \label{eq:half-step}\\
v^{n} & = & \frac{r^{n+1}-(1-c_2)r^n-c_2r^{n-1}}{2\sqrt{c_1c_3}\,\Delta{t}} + \frac{1}{4m}(\beta^n-\beta^{n+1})\, . \nonumber \\ \label{eq:on-site}
\end{eqnarray}\label{eq:velocities}\noindent
\end{subequations}
The statistical definitions of what it means to be half-step and on-site are given in Ref.~\cite{GJ}. The functional parameter $c_2=c_2(\frac{\alpha\Delta{t}}{m})$ is the one-time-step velocity attenuation, which is a function of $\alpha\Delta{t}/m$. The two other functional parameters, $c_1$ and $c_3$, are given by
\begin{subequations}
\begin{eqnarray}
2c_1 & = & 1+c_2\label{eq:c1}\\
\frac{\alpha\Delta{t}}{m}\,c_3 & = & 1-c_2 \, . \label{eq:c3}
\end{eqnarray}\label{eq:c13}\noindent
\end{subequations}
It is the choice of the function $c_2$ that distinguishes the GJ methods from each other (see Ref.~\cite{GJ} for several choices of $c_2$). The basic requirement for $c_2$ is that $c_2\rightarrow1-\frac{\alpha\Delta{t}}{m}$ for $\frac{\alpha\Delta{t}}{m}\rightarrow0$, ensuring that also $c_1\rightarrow1$ and $c_3\rightarrow1$ for $\frac{\alpha\Delta{t}}{m}\rightarrow0$ in accordance with the frictionless Verlet method \cite{Verlet,Swope,Beeman,Buneman,Hockney}. Finally, the discrete-time stochastic noise variable is defined by
\begin{eqnarray}
\beta^{n+1} & = & \int_{t_n}^{t_{n+1}}\beta(t)\,dt \, , \label{eq:beta_n}
\end{eqnarray}
such that the discrete-time fluctuation-dissipation theorem follows from Eqs.~(\ref{eq:beta_n}) and (\ref{eq:FD}) to become
\begin{subequations}
\begin{eqnarray}
\langle\beta^n\rangle & = & 0 \label{eq:FD_d_1}\\
\langle\beta^n\beta^\ell\rangle & = & 2\alpha\Delta{t}\,k_BT\,\delta_{n,\ell}\, , \label{eq:FD_d_2}
\end{eqnarray}\label{eq:FD_d}\noindent
\end{subequations}
where $\delta_{n,\ell}$ is Kronecker's delta function.

The GJ methods represented in Eq.~(\ref{eq:GJ_combo}) can equally well be written in the velocity-explicit and leap-frog Verlet forms \cite{GJ} or they can be written in the so-called splitting form that partitions the evolution into inertial, interactive, and thermodynamic operations. One way to accomplish this is given in Ref.~\cite{Josh_3}, where the pivotal pair of thermodynamic operations are condensed into one operation, and does therefore not explicitly include the half-step velocity. Here we write a GJ splitting form as follows:
\begin{subequations}
\begin{eqnarray}
v^{n+\frac{1}{4}} & = & v^n + \sqrt{\frac{c_3}{c_1}}\frac{\Delta{t}}{2m}\,f^n \label{eq:GJ_split_interact_1}\\
r^{n+\frac{1}{2}} & = & r^n + \sqrt{\frac{c_3}{c_1}}\frac{\Delta{t}}{2}\,v^{n+\frac{1}{4}} \label{eq:GJ_split_inertia_1}\\
u^{n+\frac{1}{2}} & = & \sqrt{c_1}\,v^{n+\frac{1}{4}} + \frac{1}{\sqrt{2}}\sqrt{(1-c_2)\frac{k_BT}{m}}\,\sigma^{n+1} \nonumber \\
& = & \sqrt{c_1}\,v^{n+\frac{1}{4}} + \frac{\sqrt{c_3}}{2m}\,\beta^{n+1}\label{eq:GJ_split_thermo_1}\\
v^{n+\frac{3}{4}} & = & \frac{c_2}{\sqrt{c_1}}\,u^{n+\frac{1}{2}} + \frac{1}{\sqrt{2c_1}}\sqrt{(1-c_2)\frac{k_BT}{m}}\,\sigma^{n+1} \nonumber \\
& = & \frac{c_2}{\sqrt{c_1}}\,u^{n+\frac{1}{2}} + \sqrt{\frac{c_3}{c_1}}\frac{1}{2m}\,\beta^{n+1}\label{eq:GJ_split_thermo_2}\\
r^{n+1} & = & r^{n+\frac{1}{2}} + \sqrt{\frac{c_3}{c_1}}\frac{\Delta{t}}{2}\,v^{n+\frac{3}{4}} \label{eq:GJ_split_inertia_2}\\
v^{n+1} & = & v^{n+\frac{3}{4}} + \sqrt{\frac{c_3}{c_1}}\frac{\Delta{t}}{2m}\,f^{n+1}\, , \label{eq:GJ_split_ineract_2}
\end{eqnarray}\label{eq:GJ_split}\noindent
\end{subequations}
where $\sigma^n$ is a stochastic variable with $\langle\sigma^n\rangle=0$ and $\langle\sigma^n\sigma^\ell\rangle=\delta_{n,\ell}$, and where the half-step velocity $u^{n+\frac{1}{2}}$ of Eq.~(\ref{eq:half-step}) appears explicitly in Eq.~(\ref{eq:GJ_split_thermo_1}).

Eliminating the velocities from either Eq.~(\ref{eq:GJ_combo}) or Eq.~(\ref{eq:GJ_split}) yields the purely configurational GJ stochastic Verlet form
\begin{eqnarray}
r^{n+1} & = & 2c_1\,r^n-c_2\,r^{n-1}+\frac{c_3\Delta{t}^2}{m}\,f^n+\frac{c_3\Delta{t}}{2m}(\beta^n+\beta^{n+1}) \, , \nonumber \\
 \label{eq:GJ_c}
\end{eqnarray}
which is the only stochastic Verlet form that {\bf 1)} for a harmonic potential $E_p(r)$ with a Hooke's force $f=-\kappa r$ ($\kappa>0$) and Gaussian fluctuations $\beta^n$, will yield the correct Boltzmann distribution $\rho_{c,e}(r^n)$ given by Eq.~(\ref{eq:BBoltz}), such that
\begin{subequations}
\begin{eqnarray}
\rho_{c,e}(r^n) & = & \sqrt{\frac{\kappa}{2\pi k_BT}}\exp\left(-\frac{\frac{1}{2}\kappa (r^n)^2}{k_BT}\right) \label{eq:Boltzmann_c_exact}\\
\langle r^nr^n\rangle & = & \frac{k_BT}{\kappa}\, ,
\end{eqnarray}
\end{subequations}
for any time step within the stability range ($\Omega_0\Delta{t}<2\sqrt{c_1/c_3}$); {\bf 2)} for a linear potential where $f={\rm const}$, yields the correct configurational drift velocity
\begin{eqnarray}
v_d \; = \; \left\langle\frac{r^{n+1}-r^n}{\Delta{t}}\right\rangle & = & \frac{f}{\alpha} \, ; \label{eq:v_d}
\end{eqnarray}
and {\bf 3)} for $f=0$, yields the correct Einstein diffusion given by Eq.~(\ref{eq:Einstein_D}), such that
\begin{eqnarray}
D_E & = & \lim_{n\rightarrow\infty}\frac{\left\langle(r^n-r^0)^2\right\rangle}{2d\,n\Delta{t}} \; = \;  \frac{k_BT}{\alpha}\, . \label{eq:E_D}
\end{eqnarray}
Similarly, it was found \cite{GJ} that for a single Gaussian noise value $\beta^n$ per time step, a Hooke's force $f=-\kappa r$ ($\kappa\ge0$) results in the unique half-step velocity $u^{n+\frac{1}{2}}$ in Eq.~(\ref{eq:half-step}) with correct velocity distribution function $\rho_{k,e}(u^{n+\frac{1}{2}})$ from Eq.~(\ref{eq:MBoltzmann_exact}) such that
\begin{eqnarray}
\langle u^{n+\frac{1}{2}}u^{n+\frac{1}{2}}\rangle & = & \frac{k_BT}{m}\, . \label{eq:Ek_half-step}
\end{eqnarray}
In contrast, it was also found in Ref.~\cite{GJ} that it is not possible to define an onsite velocity $v^n$ with the same time-step independent property. Given the statistical invariance of these equations upon the time step when using Gaussian noise, we will now investigate the statistical properties of $r^n$ and $u^{n+\frac{1}{2}}$ when using other distributions than Gaussian for $\beta^n$.

\section{Moment Analysis for Linear Systems}
\label{sec:moments}
We will here investigate the relevant moments of the configurational coordinate $r^n$ and half-step velocity $u^{n+\frac{1}{2}}$ as a function of the applied noise distribution $\rho(\beta)$ of $\beta^n$, which satisfies Eq.~(\ref{eq:FD_d}). For this analysis we will mostly assume a linear system with a harmonic potential
\begin{eqnarray}
E_p(r) & = & \frac{1}{2}\kappa \, r^2 \label{eq:Ep_Hooke}\\
\Rightarrow \; \; f & = & -\kappa r\, ,  \label{eq:f_Hooke}
\end{eqnarray}
such that the linearized configurational stochastic GJ methods in Eq.~(\ref{eq:GJ_c}) can be written
\begin{eqnarray}
r^{n+1} & = & 2c_1X\,r^n-c_2\,r^{n-1}+\frac{c_3\Delta{t}}{2m}(\beta^n+\beta^{n+1}) \, , \label{eq:GJ_c_lin}
\end{eqnarray}
with
\begin{eqnarray}
X & = & 1-\frac{c_3}{c_1}\frac{\Omega_0^2\Delta{t}^2}{2}\, . \label{eq:X}
\end{eqnarray}
The natural frequency $\Omega_0$ of the oscillator is given by $\Omega_0^2=\kappa/m$. 

\subsection{First and Second Moments}
\label{sec:M1M2}
\noindent
{\underline{For the first moments}} of the variables, it is obvious that Eq.~(\ref{eq:FD_d_1}) ensures both $\langle r^n\rangle=0$ (from Eq.~(\ref{eq:GJ_c_lin})) and subsequently $\langle u^{n+\frac{1}{2}}\rangle=0$ (from Eq.~(\ref{eq:half-step})) regardless of which distribution is chosen for $\beta^n$. It similarly follows that Eq.~(\ref{eq:FD_d_1}) also ensures that the drift velocity in Eq.~(\ref{eq:v_d}) for $f={\rm const}$ remains constant and correct regardless of both time step and distribution $\beta^n$.

\noindent
{\underline{For the second moments}}, $\langle r^nr^n\rangle$ and $\langle u^{n+\frac{1}{2}}u^{n+\frac{1}{2}}\rangle$, we first find $\langle r^nr^n\rangle$ as the solution to the system of equations given by the statistical averages of Eq.~(\ref{eq:GJ_c_lin}) multiplied by, respectively $r^{n-1}$, $r^n$, and $r^{n+1}$. This linear system of equations can be conveniently written
\begin{widetext}
\begin{eqnarray}
\left(\begin{array}{ccc}
1 & -2c_1X & c_2\\
0 & 2c_1 & -2c_1X\\
c_2 & -2c_1X & 1\end{array}\right)
\left(\begin{array}{c}
\langle r^{n-1}r^{n+1}\rangle\\
\langle r^{n}r^{n+1}\rangle\\
\langle r^{n}r^{n}\rangle
\end{array}\right) & = &
\left(\begin{array}{c}
0\\
1\\
2c_1X+2
\end{array}\right)\,\left(\frac{c_3\Delta{t}}{2m}\right)^2\,\langle\beta^n\beta^n\rangle\, .  \label{eq:M2_eqs}
\end{eqnarray}
\end{widetext}
We immediately observe that this equation is unaffected by the details of the distribution of $\beta^n$ for as long as Eq.~(\ref{eq:FD_d}) is satisfied. Thus, also the second moment, $\langle r^nr^n\rangle$, is unaffected by the chosen distribution of $\beta^n$, regardless of the time step $\Delta{t}$.
The corresponding second moment of the half-step velocity in Eq.~(\ref{eq:half-step}) is then given by
\begin{eqnarray}
\langle u^{n+\frac{1}{2}}u^{n+\frac{1}{2}}\rangle & = & \frac{2}{c_3\,\Delta{t}^2}(\langle r^nr^n\rangle-\langle r^nr^{n+1}\rangle)\, ,  \label{eq:M2_u}
\end{eqnarray}
where both correlations on the right hand side are solutions found from Eq.~(\ref{eq:M2_eqs}), which is unchanged for as long as Eq.~(\ref{eq:FD_d_2}) is satisfied. Thus, the second moment of the half-step velocity is also unaffected by the chosen distribution of $\beta^n$, regardless of the time step $\Delta{t}$. The averages of potential (Eq.~(\ref{eq:Ep_Hooke})) and kinetic energies,
\begin{eqnarray}
\left\langle E_p(r^n)\right\rangle & = & \frac{\kappa}{2}\langle r^nr^n\rangle \; = \; \frac{1}{2}k_BT \label{eq:Ep_ave}\\
\left\langle E_k(u^{n+\frac{1}{2}})\right\rangle & = &  \frac{1}{2}\frac{m}{c_3\,\Delta{t}^2}(\langle r^nr^n\rangle-\langle r^nr^{n+1}\rangle)\nonumber \\
& = & \frac{1}{2}k_BT \, ,\label{eq:Ek_ave}
\end{eqnarray}
are therefore unaffected for any applied stochastic variable $\beta^n$ satisfying Eq.~(\ref{eq:FD_d}).

Since the second moments of the coordinates $r^n$ and $u^{n+\frac{1}{2}}$ of the GJ methods are invariant to the specific choice of noise distribution that satisfies Eq.~(\ref{eq:FD_d}), we conclude that the GJ methods display time-step invariance in both potential and kinetic energy, and therefore temperature, regardless of the applied $\beta^n$ distribution. The invariance also extends to the above-mentioned diffusion Eqs.~(\ref{eq:Einstein_D}) and (\ref{eq:E_D}), since this quantity is rooted in only second moments of the noise variable.

Numerical validation of the time-step independence of the second moments is shown in Figs.~\ref{fig:fig_2}ab and \ref{fig:fig_6}ab for a couple of non-Gaussian noise variables, and simulations are discussed in Sec.~\ref{sec:SimDisc}. While the invariance of the second moments against the choice of $\beta^n$ distribution is appealing, as it gives the impression that temperature is correctly achieved, it may not be sufficient to ensure proper thermodynamics sampling as we will see in the following.

\subsection{Third Moments}
\label{sec:M3}
The third moments of $r^n$ and $u^{n+\frac{1}{2}}$ are generally not needed for the calculations of thermodynamic quantities, especially if the system potentials are symmetric. However, they may be relevant to study in case asymmetric fluctuations are applied to the system for some reason. Following the same procedure that was used for calculating the second moments, we form the statistical averages of the product of Eq.~(\ref{eq:GJ_c_lin}) with the six unique combinations of $r^{\ell_1}r^{\ell_2}$, where $\ell_i=n-1,n,n+1$. The result is the linear system
\begin{widetext}
\begin{eqnarray}
{\cal A}_3	\left(\begin{array}{c}
	\langle r^{n-1}r^{n-1}r^{n+1}\rangle \\
	\langle r^{n-1}r^{n}r^{n+1}\rangle \\
	\langle r^{n-1}r^{n+1}r^{n+1}\rangle \\
	\langle r^{n}r^{n}r^{n+1}\rangle \\
	\langle r^{n}r^{n+1}r^{n+1}\rangle \\
	\langle r^{n}r^{n}r^{n}\rangle
	\end{array}\right) & = & \left(\frac{k_BT}{\kappa}\right)^{\frac{3}{2}}\,{\cal R}  \; = \; 	\left(\begin{array}{c}
	\langle r^{n-1}r^{n-1}\beta^n\rangle\ \\
	\langle r^{n-1}r^{n}\beta^n\rangle\\
	\langle r^{n-1}r^{n+1}(\beta^n+\beta^{n+1})\rangle\\
	\langle r^{n  }r^{n  }\beta^n\rangle\\
	\langle r^{n  }r^{n+1}(\beta^n+\beta^{n+1})\rangle\\
	\langle r^{n+1}r^{n+1}(\beta^n+\beta^{n+1})\rangle
	\end{array}\right)\frac{c_3\Delta{t}}{2m}\, ,  \label{eq:Ar3R}
\end{eqnarray}
\end{widetext}
where
\begin{eqnarray}
{\cal A}_3 & = & \left(\begin{array}{cccccc}
1	&	0 &	0 &-2c_1X	&	0  &c_2	\\
0	& 1	 & 0	& c_2 &-2c_1X	& 0 	\\
c_2	&-2c_1X	& 1	&	0 &	0 &	0 \\
0 	&	0 & 	0 &1	&c_2	&-2c_1X \\
0 	&c_2	& 	0 &-2c_1X	&1	&	0 \\
0 	&	0 &c_2 	&	0 &-2c_1X	& 1		
	\end{array}\right)\, . \label{eq:A}
\end{eqnarray}
The right hand side ${\cal R} = \{{\cal R}_i\}$ can after some calculations be expressed as
\begin{subequations}
\begin{eqnarray}
{\cal R}_1 & = & 0 \label{eq:R_1}\\
{\cal R}_2 & = & 0 \label{eq:R_2}\\
{\cal R}_3 & = & 0 \label{eq:R_3}\\
{\cal R}_4 & = &
 (1-c_2)^3\left(\frac{\kappa\Delta{t}}{\alpha}\right)^\frac{3}{2}\,\Gamma_3\label{eq:R_4}\\
{\cal R}_5 & = & (2c_1X+1){\cal R}_4 \label{eq:R_5}\\
{\cal R}_6 & = & (2c_1X(2c_1X+2)+2){\cal R}_4\, ,  \label{eq:R_6}
\end{eqnarray}\label{eq:R3}\noindent
\end{subequations}
where the fluctuations $\beta^n$ satisfy Eq.~(\ref{eq:FD_d}), and the skewness is denoted by $\Gamma_3$,
\begin{eqnarray}
\Gamma_3 & = & \frac{\left\langle\beta^n\beta^n\beta^n\right\rangle}{\left\langle\beta^n\beta^n\right\rangle^\frac{3}{2}}\, . 
\end{eqnarray}
As is the case when using any symmetric distribution for $\beta^n$, all odd moments for the configurational coordinate $r^n$ will be zero for a Gaussian distribution of $\beta^n$ applied to a symmetric potential, $E_p(r)$.

The third moment of the velocity $u^{n+\frac{1}{2}}$ is given by
\begin{eqnarray}
&& \left\langle u^{n+\frac{1}{2}}u^{n+\frac{1}{2}}u^{n+\frac{1}{2}}\right\rangle \; = \; \frac{1}{\left(\Delta{t}\,\sqrt{c_3}\right)^3}\left\langle(r^{n+1}-r^n)^3\right\rangle \nonumber \\
& & = \;  \frac{3}{\left(\Delta{t}\,\sqrt{c_3}\right)^3}\left(\left\langle r^nr^nr^{n+1}\right\rangle-\left\langle r^nr^{n+1}r^{n+1}\right\rangle\right)\, ,  \label{eq:uuu}
\end{eqnarray}
where the two correlations on the right hand side are part of the solution to Eq.~(\ref{eq:Ar3R}).

For any given choice of GJ method ($c_2$) the above system of equations can be easily solved numerically for given system parameters and skewness $\Gamma_3$. Examples of this will be discussed in Sec.~\ref{sec:SimDisc}, and are shown in Fig.~\ref{fig:fig_6}cd along with the results of corresponding numerical solution of Eqs.~(\ref{eq:GJ_combo}) and (\ref{eq:f_Hooke}) for a particular asymmetric noise distribution. Closed form expressions are obtained in the following for a couple of special cases.

\begin{description}
\item[{\underline{Special Case, $X=0$}}] The matrix ${\cal A}_3$ and the vector ${\cal R}$ simplify considerably with easy accessible solutions
\begin{subequations}
\begin{eqnarray}
\langle r^nr^nr^n\rangle & = &  \frac{2}{1+c_2^3}\left(\frac{k_BT}{\kappa}\right)^{\frac{3}{2}}\,{\cal R}_4 \label{eq:r0r0r0_X0}\\
\langle r^nr^nr^{n+1}\rangle & = &  \frac{1-c_2}{1+c_2^3}\left(\frac{k_BT}{\kappa}\right)^{\frac{3}{2}}\,{\cal R}_4  \label{eq:r0r0r1_X0}\\
\langle r^nr^{n+1}r^{n+1}\rangle & = & \frac{1+c_2^2}{1+c_2^3}\left(\frac{k_BT}{\kappa}\right)^{\frac{3}{2}}\,{\cal R}_4 \, ,  \label{eq:r0r1r1_X0}
\end{eqnarray}
\end{subequations}
where we have used that this special case for $\Omega_0^2\Delta{t}^2=2\frac{c_3}{c_1}$ offers ${\cal R}_6=2{\cal R}_5=2{\cal R}_4$, and where
\begin{eqnarray}
{\cal R}_4 & = & \left(\frac{1-c_2^2}{2}\right)^{\frac{3}{2}}\,\Gamma_3\, . 
\end{eqnarray}
Equations (\ref{eq:uuu}), (\ref{eq:r0r0r1_X0}), and (\ref{eq:r0r1r1_X0}) give
\begin{eqnarray}
&&\left\langle \left(u^{n+\frac{1}{2}}\right)^3\right\rangle \; = \; - \frac{3}{\left(\Delta{t}\sqrt{c_3}\right)^3}\frac{c_2}{c_2^2-c_2+1}\left(\frac{k_BT}{\kappa}\right)^{\frac{3}{2}}{\cal R}_4 \nonumber \\ \\
&& = \; -\frac{3}{2\sqrt{2}}\,\left(\frac{1-c_2^2}{1+c_2}\right)^\frac{3}{2}\frac{c_2}{c_2^2-c_2+1}\,\left(\frac{k_BT}{m}\right)^\frac{3}{2} \,\Gamma_3 \, . \label{eq:uuu_X0}
\end{eqnarray}
Examples of the use of these expressions are shown in Fig.~\ref{fig:fig_6}cd as open markers ($\circ$) for a particular asymmetric noise distribution.
\item[{\underline{Special Case, $c_2=0$}}] In this case we have that $c_1=\frac{1}{2}$, $c_3 = m/\alpha\Delta{t}$, and $X=1-\Delta{t}\kappa/\alpha$. The simplified matrix ${\cal A}_3$ and vector ${\cal R}$ yield
\begin{subequations}
\begin{eqnarray}
\langle r^nr^nr^n\rangle & = & \frac{3X^2+3X+2}{1-X^3}\left(\frac{k_BT}{\kappa}\right)^{\frac{3}{2}}\,{\cal R}_4  \label{eq:r0r0r0_c20}\\
\langle r^nr^nr^{n+1}\rangle & = & \frac{2X^3+3X^2+2X+1}{1-X^3}\left(\frac{k_BT}{\kappa}\right)^{\frac{3}{2}}{\cal R}_4 \label{eq:r0r0rp_c20}\\
\langle r^nr^{n+1}r^{n+1}\rangle & = & \frac{X^4+2X^3+2X^2+2X+1}{1-X^3}\left(\frac{k_BT}{\kappa}\right)^{\frac{3}{2}}{\cal R}_4\, , \nonumber \\  \label{eq:r0rprp_c20}
\end{eqnarray}
\end{subequations}
where
\begin{eqnarray}
{\cal R}_4 & = & \left(\frac{\kappa\Delta{t}}{2\alpha}\right)^\frac{3}{2}\,\Gamma_3\, .
\end{eqnarray}
Equations (\ref{eq:uuu}), (\ref{eq:r0r0rp_c20}), and (\ref{eq:r0rprp_c20}) give
\begin{eqnarray}
\left\langle \left(u^{n+\frac{1}{2}}\right)^3\right\rangle & = & \frac{3X^2}{2\sqrt{2}}\frac{1-X^2}{1-X^3}\left(\frac{k_BT}{m}\right)^\frac{3}{2}\Gamma_3\, . \label{eq:uuu_c20}
\end{eqnarray}
Examples of the use of these expressions are shown in Fig.~\ref{fig:fig_6}cd as closed markers ($\bullet$) for a particular asymmetric noise distribution.
\end{description}

\subsection{Fourth Moments}
\label{sec:M4}
The fourth moments of $r^n$ and $u^{n+\frac{1}{2}}$ are needed for the calculations of energy and pressure fluctuations, which are the basis of several important thermodynamic quantities, such as heat capacity, thermal expansion, compressibility, and the thermal pressure coefficient \cite{AllenTildesley}. Fourth moments are therefore critical for proper thermodynamic characterization of a system. Following the same procedure that was used for calculating the second and third moments, we form the statistical averages of the product of Eq.~(\ref{eq:GJ_c_lin}) with the ten unique combinations of $r^{\ell_1}r^{\ell_2}r^{\ell_3}$, where $\ell_i=n-1,n,n+1$. The result is the linear system
\begin{widetext}
\begin{eqnarray}
{\cal A}_4	\left(\begin{array}{c}
	\langle r^{n-1}r^{n-1}r^{n-1}r^{n+1}\rangle \\
	\langle r^{n-1}r^{n-1}r^{n}r^{n+1}\rangle \\
	\langle r^{n-1}r^{n-1}r^{n+1}r^{n+1}\rangle \\
	\langle r^{n-1}r^{n}r^{n}r^{n+1}\rangle \\
	\langle r^{n-1}r^{n}r^{n+1}r^{n+1}\rangle \\
	\langle r^{n-1}r^{n+1}r^{n+1}r^{n+1}\rangle \\
	\langle r^{n}r^{n}r^{n}r^{n+1}\rangle \\
	\langle r^{n}r^{n}r^{n+1}r^{n+1}\rangle \\
	\langle r^{n}r^{n+1}r^{n+1}r^{n+1}\rangle \\
	\langle r^{n}r^{n}r^{n}r^{n}\rangle 
	\end{array}\right) & = &  \left(\frac{k_BT}{\kappa}\right)^2 {\cal R}  \; = \; 	\left(\begin{array}{c}
	0 \\
	\langle r^{n-1}r^{n-1}r^{n}\beta^n\rangle\\
	\langle r^{n-1}r^{n-1}r^{n+1}(\beta^n+\beta^{n+1})\rangle\\
	\langle r^{n-1}r^{n  }r^{n  }\beta^n\rangle\\
	\langle r^{n-1}r^{n  }r^{n+1}(\beta^n+\beta^{n+1})\rangle\\
	\langle r^{n-1}r^{n+1}r^{n+1}(\beta^n+\beta^{n+1})\rangle\\
	\langle r^{n  }r^{n  }r^{n  }\beta^n\rangle\\
	\langle r^{n  }r^{n  }r^{n+1}(\beta^n+\beta^{n+1})\rangle\\
	\langle r^{n  }r^{n+1}r^{n+1}(\beta^n+\beta^{n+1})\rangle\\
	\langle r^{n+1}r^{n+1}r^{n+1}(\beta^n+\beta^{n+1})\rangle
	\end{array}\right)\frac{c_3\Delta{t}}{2m} \, , \label{eq:Ar4R}
\end{eqnarray}
where
\begin{eqnarray}
{\cal A}_4 & = & \left(\begin{array}{cccccccccc}
1	&	0 &	0 &	0 &	0 &	0 &-2c_1X	&	0 &	0 &c_2	\\
0	& 1	&	0 &	0 &	0 &	0 &c_2	&-2c_1X	& 0 &0	\\
c_2	&-2c_1X	& 1	&	0 &	0 &	0 &	0 &	0 &	0 &0 	\\
0 	&	0 & 	0 &1	&	0 &	0 &0 	&c_2	&-2c_1X	&	0 \\
0 	&c_2	& 	0 &-2c_1X	&1	&	0 &	0 &	0 &	0 &	0 \\
0 	&	0 &c_2 	&	0 &-2c_1X	&1	&	0 &	0 &0 	&	0 \\
0 	&	0 & 	0 &	0 &	0 &	0 &1	&	0 &c_2	&-2c_1X	\\
0 	&	0 & 	0 &c_2	&	0 &	0 &-2c_1X	&1	&	0 &0 	\\
0 	&	0 & 	0 &	0 &c_2	&	0 &	0 &-2c_1X	&1	&0 	\\
0 	&	0 & 	0 &	0 &	0 &c_2	&	0 &	0 &-2c_1X	&1	
	\end{array}\right)\, . \label{eq:A}
\end{eqnarray}
The right hand side ${\cal R} = \{{\cal R}_i\}$ can after some cumbersome work be expressed as
\begin{subequations}
\begin{eqnarray}
{\cal R}_1 & = & 0 \label{eq:R_1}\\
{\cal R}_2 & = & \frac{1}{2}(1-c_2^2)(1-X)\label{eq:R_2}\\
{\cal R}_3 & = & [2+2c_1X]{\cal R}_2 \label{eq:R_3}\\
{\cal R}_4 & = &  [2c_1X+1-c_2]{\cal R}_2 \label{eq:R_4}\\
{\cal R}_5 & = &  [2c_1X(2c_1X+2-c_2)+1-2c_2]{\cal R}_2 \label{eq:R_5} \\
{\cal R}_6 & = &  [(2c_1X)^3+(2c_1X)^2(3-c_2)+4c_1X(1-2c_2)-4c_2]{\cal R}_2 \label{eq:R_6}\\
{\cal R}_7 & = &  [3+(\Gamma_4-3){\cal R}_2]{\cal R}_2 \label{eq:R_7}\\
{\cal R}_8 & = &  [4c_1X+3-c_2+(2c_1X+1)(\Gamma_4-3){\cal R}_2]{\cal R}_2 \label{eq:R_8}\\
{\cal R}_9 & = &  [(2c_1X)^2+2c_1X(3-c_2)+3-2c_2+(1+2c_1X)^2(\Gamma_4-3){\cal R}_2]{\cal R}_2 \label{eq:R_9}\\
{\cal R}_{10} & = &  [3(2c_1X+2)+(1+(1+2c_1X)^3)(\Gamma_4-3){\cal R}_2]{\cal R}_2\, ,  \label{eq:R_10}
\end{eqnarray}\label{eq:R}
\end{subequations}
\end{widetext}
where we have used Eq.~(\ref{eq:FD_d_2}). Notice that ${\cal R}_7$-${\cal R}_{10}$ in Eqs.~(\ref{eq:R_7})-(\ref{eq:R_10}) depend on the kurtosis $\Gamma_4$
\begin{eqnarray}
\Gamma_4 & = & \frac{\langle\beta^n\beta^n\beta^n\beta^n\rangle}{\langle\beta^n\beta^n\rangle^2}\, , \label{eq:kurtosis}
\end{eqnarray}
 which is not determined by the fluctuation-dissipation relationship in Eq.~(\ref{eq:FD_d}).

While Eqs.~(\ref{eq:Ar4R})-(\ref{eq:R}) are not immediately informative, they provide a straightforward path to calculating the sought-after fluctuation $\sigma_p$ in potential energy of Eq.~(\ref{eq:Ep_Hooke})
\begin{eqnarray}
\sigma_p & = & \sqrt{\left\langle E^2_p(r^n)\right\rangle-\left\langle E_p(r^n)\right\rangle^2} \nonumber \\
& = & \frac{\kappa}{2}\sqrt{\langle r^nr^nr^nr^n\rangle-\langle r^nr^n\rangle^2}\, .  \label{eq:sigma_p}
\end{eqnarray}
Equations (\ref{eq:Ar4R})-(\ref{eq:R}) also lead directly to the fluctuations $\sigma_k$ in kinetic energy of the velocity from Eq.~(\ref{eq:half-step})
\begin{eqnarray}
\sigma_k & = & \sqrt{\left\langle E^2_k(u^{n+\frac{1}{2}})\right\rangle-\left\langle E_k(u^{n+\frac{1}{2}})\right\rangle^2} \nonumber \\
& = & \frac{m}{2}\sqrt{\langle u^{n+\frac{1}{2}}u^{n+\frac{1}{2}}u^{n+\frac{1}{2}}u^{n+\frac{1}{2}}\rangle-\langle u^{n+\frac{1}{2}}u^{n+\frac{1}{2}}\rangle^2}  \, , \nonumber \\ \label{eq:sigma_k}
\end{eqnarray}
where $\langle u^{n+\frac{1}{2}}u^{n+\frac{1}{2}}\rangle$ is given by Eq.~(\ref{eq:Ek_half-step}), and
\begin{eqnarray}
&& \langle u^{n+\frac{1}{2}}u^{n+\frac{1}{2}}u^{n+\frac{1}{2}}u^{n+\frac{1}{2}}\rangle \; = \; \frac{1}{c_3^2\Delta{t}^4}\langle(r^{n+1}-r^{n})^4\rangle \nonumber \\
& & = \; \frac{2}{c_3^2\Delta{t}^4}(\langle r^nr^nr^nr^n\rangle-2\langle r^nr^nr^nr^{n+1}\rangle\nonumber \\
&& + \; 3\langle r^nr^nr^{n+1}r^{n+1}\rangle-2\langle r^nr^{n+1}r^{n+1}r^{n+1}\rangle)\, .
\end{eqnarray}
All the correlations on the right hand side are given from the solution to Eqs.~(\ref{eq:Ar4R})-(\ref{eq:R}).

\begin{figure}[t]
\centering
\scalebox{0.45}{\centering \includegraphics[trim={1.5cm 2.0cm 1cm 2.0cm},clip]{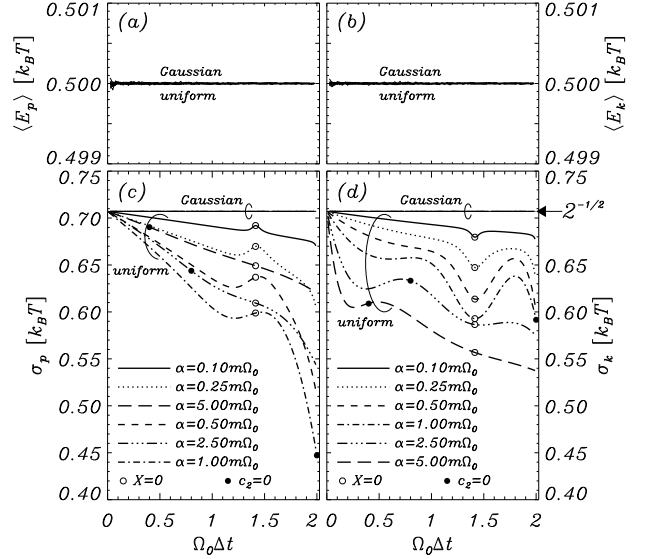}}
\caption{Simulations and analyses of average potential and kinetic energies, $\langle E_p\rangle$ from Eq.~(\ref{eq:Ep_ave}) (a) and $\langle E_k\rangle$ from Eq.~(\ref{eq:Ek_ave}) (b), and their respective fluctuations, $\sigma_p$ from Eq.~(\ref{eq:sigma_p}) (c) and $\sigma_k$ from Eq.~(\ref{eq:sigma_k}) (d), as a function of reduced time step for the harmonic system, Eqs.~(\ref{eq:GJ_combo}) and (\ref{eq:f_Hooke}), with non-zero temperature $T$. Results are shown for Gaussian $\beta_G^n$ (Eq.~(\ref{eq:Gauss})) and uniform $\beta_U^n$ (Eq.~(\ref{eq:Flatn})) noise distributions. Several friction parameters $\alpha$ are used, as indicated in the figure. Simulation results are indistinguishable from the analytical ones. Markers indicate special cases for $c_2=0$ ($\bullet$ from Eqs.~(\ref{eq:c20_sc2})-(\ref{eq:c20_sk2})) and $X=0$ ($\circ$ from Eqs.~(\ref{eq:X0sp}) and (\ref{eq:X0sk})).
}
\label{fig:fig_2}
\end{figure}

\begin{figure}[t]
\centering
\scalebox{0.45}{\centering \includegraphics[trim={1.5cm 2.0cm 1cm 2.0cm},clip]{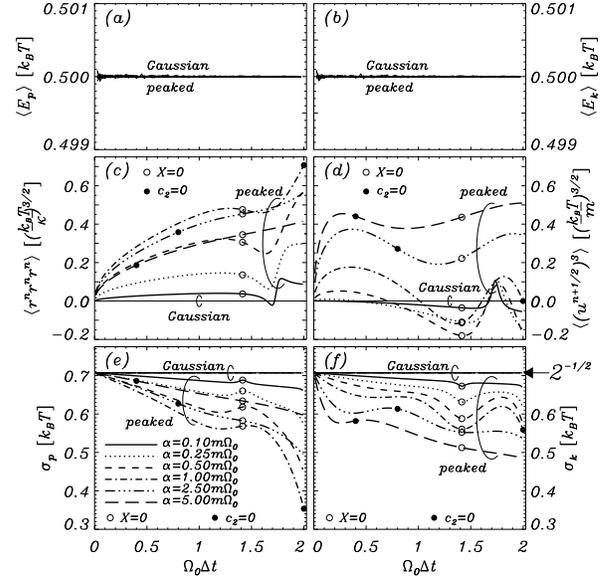}}
\caption{Simulations and analyses of average potential and kinetic energies, $\langle E_p\rangle$ from Eq.~(\ref{eq:Ep_ave}) (a) and $\langle E_k\rangle$ from Eq.~(\ref{eq:Ek_ave}) (b), the respective third moments $\langle(r^n)^3\rangle$ from Eqs.~(\ref{eq:Ar3R})-(\ref{eq:R3}) (c) and $\langle(u^{n+\frac{1}{2}})^3\rangle$ from Eq.~(\ref{eq:uuu}) (d), and the respective energy fluctuations, $\sigma_p$ from Eq.~(\ref{eq:sigma_p}) (e) and $\sigma_k$ from Eq.~(\ref{eq:sigma_k}) (f), as a function of reduced time step for the harmonic system, Eqs.~(\ref{eq:GJ_combo}) and (\ref{eq:f_Hooke}), with non-zero temperature $T$. Results are shown for Gaussian $\beta_G^n$ (Eq.~(\ref{eq:Gauss})) and the asymmetrically peaked $\beta_P^n$ (Eq.~(\ref{eq:Peakn})) noise distributions. Several friction parameters $\alpha$ are used, as indicated in the figure. Simulation results are indistinguishable from the analytical ones. Markers indicate special cases for $c_2=0$ ($\bullet$ from Eqs.~(\ref{eq:r0r0r0_c20}) and (\ref{eq:uuu_c20}), Eqs.~(\ref{eq:c20_sc2})-(\ref{eq:c20_sk2})) and $X=0$ ($\circ$ from Eqs.~(\ref{eq:r0r0r0_X0}) and (\ref{eq:uuu_X0}), Eqs.~(\ref{eq:X0sp}) and (\ref{eq:X0sk})).
}
\label{fig:fig_6}
\end{figure}

For any given choice of GJ method ($c_2$) the above system of equations can be easily solved numerically for given system parameters and kurtosis $\Gamma_4$. Examples of this will be discussed in Sec.~\ref{sec:SimDisc}, and are shown in Figs.~\ref{fig:fig_2}cd and \ref{fig:fig_6}ef along with the results of corresponding numerical solution of Eqs.~(\ref{eq:GJ_combo}) and (\ref{eq:f_Hooke}) for particular noise distributions. 
Closed form expressions can be given in the special cases of $X=0$ and $c_2=0$.

\begin{description}
\item[{\underline{Special Case, $X=0$}}] 
This particular time step is
\begin{eqnarray}
\Omega_0\Delta{t} & = & \sqrt{2}\sqrt{\frac{c_1}{c_3}}\, . 
\end{eqnarray}
Cumbersome, yet straightforward, algebra leads to the closed expressions for $\langle r^nr^nr^nr^n\rangle$ and $\langle u^{n+\frac{1}{2}}u^{n+\frac{1}{2}}u^{n+\frac{1}{2}}u^{n+\frac{1}{2}}\rangle$ for $X=0$; i.e., for $\Omega_0\Delta{t}=\sqrt{2}\sqrt{c1/c3}$. The corresponding fluctuations in potential and kinetic energies can be written
\begin{eqnarray}
\sigma_p & = & \frac{\kappa}{2}\sqrt{\langle r^nr^nr^nr^n\rangle - \langle r^nr^n\rangle^2} \nonumber \\
& = & \frac{k_BT}{\sqrt{2}}\, \sqrt{\frac{1}{4}\frac{1+7c_2^2}{1+c_2^2}+\frac{1}{4}\Gamma_4\frac{1-c_2^2}{1+c_2^2}} \label{eq:X0sp}\\
\sigma_k & = & \frac{m}{2}\sqrt{\langle u^{n+\frac{1}{2}}u^{n+\frac{1}{2}}u^{n+\frac{1}{2}}u^{n+\frac{1}{2}}\rangle - \langle u^{n+\frac{1}{2}}u^{n+\frac{1}{2}}\rangle^2} \nonumber \\
& = & \frac{k_BT}{\sqrt{2}}\, \sqrt{\frac{1}{2}\frac{c_2^3}{1+c_2^2}\left[3-\Gamma_4\right]+\frac{1}{4}\left[1+\Gamma_4\right] }\, . \label{eq:X0sk}
\end{eqnarray}
Notice that if $\beta^n$ is a Gaussian distributed variable, where the kurtosis is $\Gamma_4=3$, the above energy fluctuations yield the expected values, $\sigma_p=\sigma_k=k_BT/\sqrt{2}$.

Examples of the use of these expressions are shown in Figs.~\ref{fig:fig_2}cd and \ref{fig:fig_6}ef as open markers ($\circ$) for particular noise distributions.

\item[{\underline{Special Case, $c_2=0$}}] 
This special case (GJ-0, see Refs.~\cite{GJ}) corresponds to the first order difference equation
\begin{eqnarray}
r^{n+1} & = & r^n+\frac{1}{\alpha}\left[\Delta{t}\,f^n+\frac{1}{2}(\beta^{n}+\beta^{n+1})\right] \, , 
\end{eqnarray}
which can approximate the solution to the Langevin equation with no inertia; i.e., Brownian motion. The linearized, Hooke's law system ($f=-\kappa r$) reads
\begin{eqnarray}
r^{n+1} & = & Xr^n+\frac{1}{2\alpha}(\beta^n+\beta^{n+1})\, , 
\end{eqnarray}
with $X=1-\kappa\Delta{t}/\alpha$, in correspondence with Eqs.~(\ref{eq:GJ_c_lin}) and (\ref{eq:X}). This simplicity allows the last four of the ten equations in Eq.~(\ref{eq:Ar4R}) to decouple from the rest, and the fourth moments can be calculated to yield the average and fluctuations in potential energy
\begin{eqnarray}
\left\langle E_p(r^n)\right\rangle & = & \frac{\kappa}{2}\,\left\langle r^nr^n\right\rangle \; = \; \frac{1}{2}k_BT
\end{eqnarray}
\begin{widetext}
\begin{eqnarray}
\sigma_p & = & \frac{\kappa}{2}\,\sqrt{\left\langle r^nr^nr^nr^n\right\rangle-\left\langle r^nr^n\right\rangle^2} \; 
 = \; \frac{k_BT}{\sqrt{2}}\,\sqrt{\frac{\frac{1}{2}X^4\left[3-\Gamma_4\right]+\frac{1}{4}X^3\left[7-\Gamma_4\right]+\frac{1}{4}(X^2+X+1)\left[1+\Gamma_4\right]}{X^3+X^2+X+1}}\, . \label{eq:c20_sc2}
\end{eqnarray}
\end{widetext}
We also here notice that if $\beta^n$ is a Gaussian distributed variable, where the kurtosis is $\Gamma_4=3$, the potential energy fluctuations yield the expected value, $\sigma_p=k_BT/\sqrt{2}$.

Similarly, we can find the fluctuations $\sigma_k$ in kinetic energy from the correlation
\begin{widetext}
\begin{eqnarray}
&& \left\langle\left(u^{n+\frac{1}{2}}\right)^4\right\rangle \; = \; (1-X)^2\frac{4\sigma_p^2+(k_BT)^2}{m^2}+6\left(\frac{k_BT}{m}\right)^2X^2(1-X)
 -  \left(\frac{k_BT}{m}\right)^2(3+\Gamma_4)X\left(1-\frac{3}{2}X+X^2\right)\, ,  \label{eq:c20_u4}
\end{eqnarray}
\end{widetext}
such that
\begin{eqnarray}
\sigma_k & = & \frac{1}{2}m\,\sqrt{\left\langle\left(u^{n+\frac{1}{2}}\right)^4\right\rangle-\left(\frac{k_BT}{m}\right)^2}\, , \label{eq:c20_sk2}
\end{eqnarray}
where we have used Eq.~(\ref{eq:Ek_half-step}). Thus, $\sigma_k$ can be easily computed also in this special case. We again notice that for Gaussian fluctuations, where $\Gamma_4=3$, the kinetic energy fluctuations yield the expected value $\sigma_k=k_BT/\sqrt{2}$.

Examples of the use of these expressions are shown in Figs.~\ref{fig:fig_2}cd and \ref{fig:fig_6}ef as closed markers ($\bullet$) for particular noise distributions.

\end{description}

\begin{figure}[t]
\centering
\scalebox{0.4}{\centering \includegraphics[trim={1.5cm 2.0cm 1cm 3.5cm},clip]{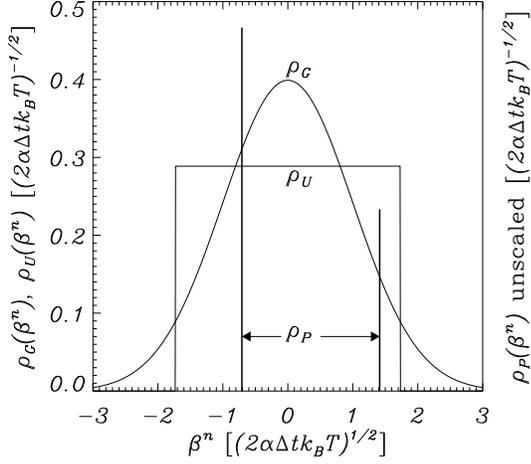}}
\caption{The three applied distributions, $\rho_G(\beta^n)$ (Gaussian), $\rho_U(\beta^n)$ (Uniform), and $\rho_P(\beta^n)$ (Peaked), given in Eqs.~(\ref{eq:Dist_Gauss}), (\ref{eq:Dist_Flatn}), and (\ref{eq:Dist_Peakn}), respectively. All three distributions satisfy the discrete-time fluctuation-dissipation relationship, Eq.~(\ref{eq:FD_d}).
}
\label{fig:fig_1}
\end{figure}

\begin{figure}[t]
\centering
\scalebox{0.45}{\centering \includegraphics[trim={1.5cm 2.0cm 1cm 1.5cm},clip]{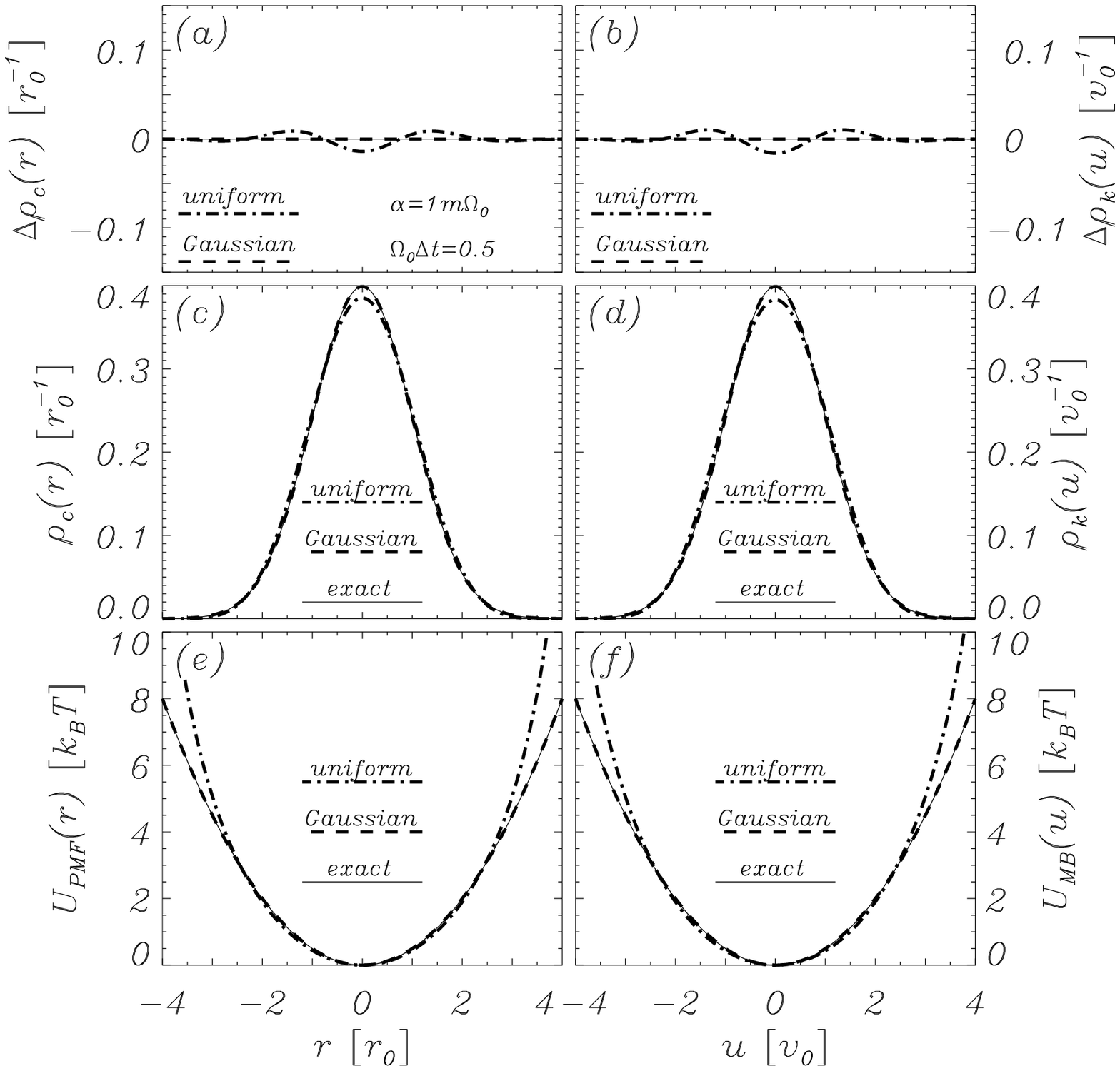}}
\caption{Configurational and kinetic distributions $\rho_c(r^n)$ and $\rho_k(u^{n+\frac{1}{2}})$ for simulations with Gaussian and uniform noise applied to the linear system Eqs.~(\ref{eq:GJ_combo}) and (\ref{eq:f_Hooke}) with $\alpha=1m\Omega_0$, $\Omega_0\Delta{t}=0.5$, $k_BT=E_0$. Shown curves are Eq.~(\ref{eq:Boltz}) (solid, labeled exact), simulations data using Gaussian noise $\beta_G^n$ (dashed, labeled Gaussian), and simulation data using uniform noise $\beta_U^n$ (dash-dot, labeled uniform). Distributions from simulations using Gaussian noise are indistinguishable from the exact references Eq.~(\ref{eq:Boltz}). (c) and (d) show the distributions, (a) and (b) show the deviations from the exact distributions Eq.~(\ref{eq:Boltz}), and (e) and (f) show the effective potential (PMF) and kinetic (MB) potentials (Eq.~(\ref{eq:PMF_MB})) derived from the distributions.
}
\label{fig:fig_3}
\end{figure}

\begin{figure}[t]
\centering
\scalebox{0.45}{\centering \includegraphics[trim={1.5cm 2.0cm 1cm 1.5cm},clip]{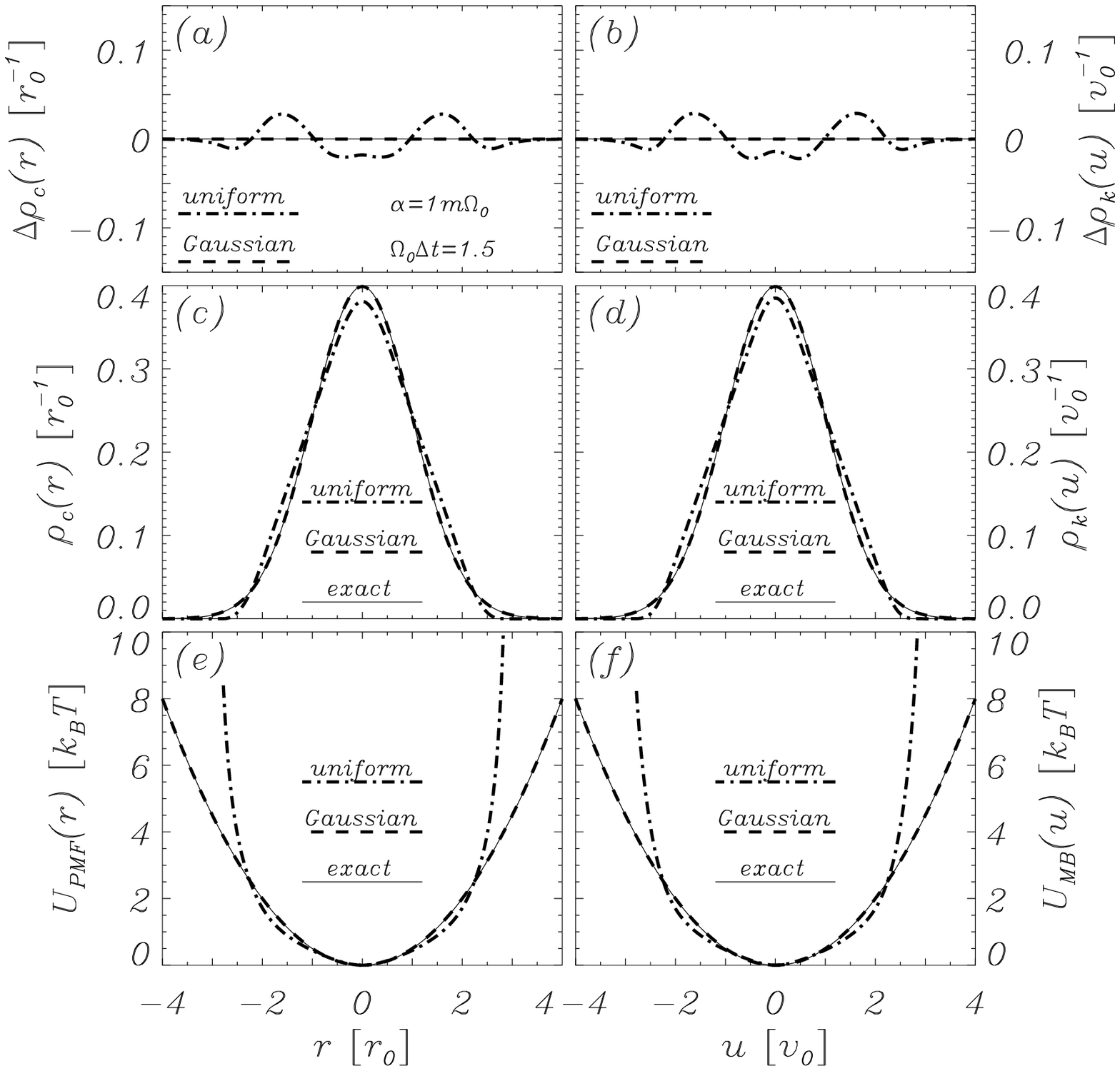}}
\caption{Configurational and kinetic distributions $\rho_c(r^n)$ and $\rho_k(u^{n+\frac{1}{2}})$ for simulations with Gaussian and uniform noise applied to the linear system Eqs.~(\ref{eq:GJ_combo}) and (\ref{eq:f_Hooke}) with $\alpha=1m\Omega_0$, $\Omega_0\Delta{t}=1.5$, $k_BT=E_0$. Shown curves are Eq.~(\ref{eq:Boltz}) (solid, labeled exact), simulations data using Gaussian noise $\beta_G^n$ (dashed, labeled Gaussian), and simulation data using uniform noise $\beta_U^n$ (dash-dot, labeled uniform). Distributions from simulations using Gaussian noise are indistinguishable from the exact references Eq.~(\ref{eq:Boltz}). (c) and (d) show the distributions, (a) and (b) show the deviations from the exact distributions Eq.~(\ref{eq:Boltz}), and (e) and (f) show the effective potential (PMF) and kinetic (MB) potentials (Eq.~(\ref{eq:PMF_MB})) derived from the distributions.
}
\label{fig:fig_4}
\end{figure}

\begin{figure}[t]
\centering
\scalebox{0.45}{\centering \includegraphics[trim={1.5cm 2.0cm 1cm 1.5cm},clip]{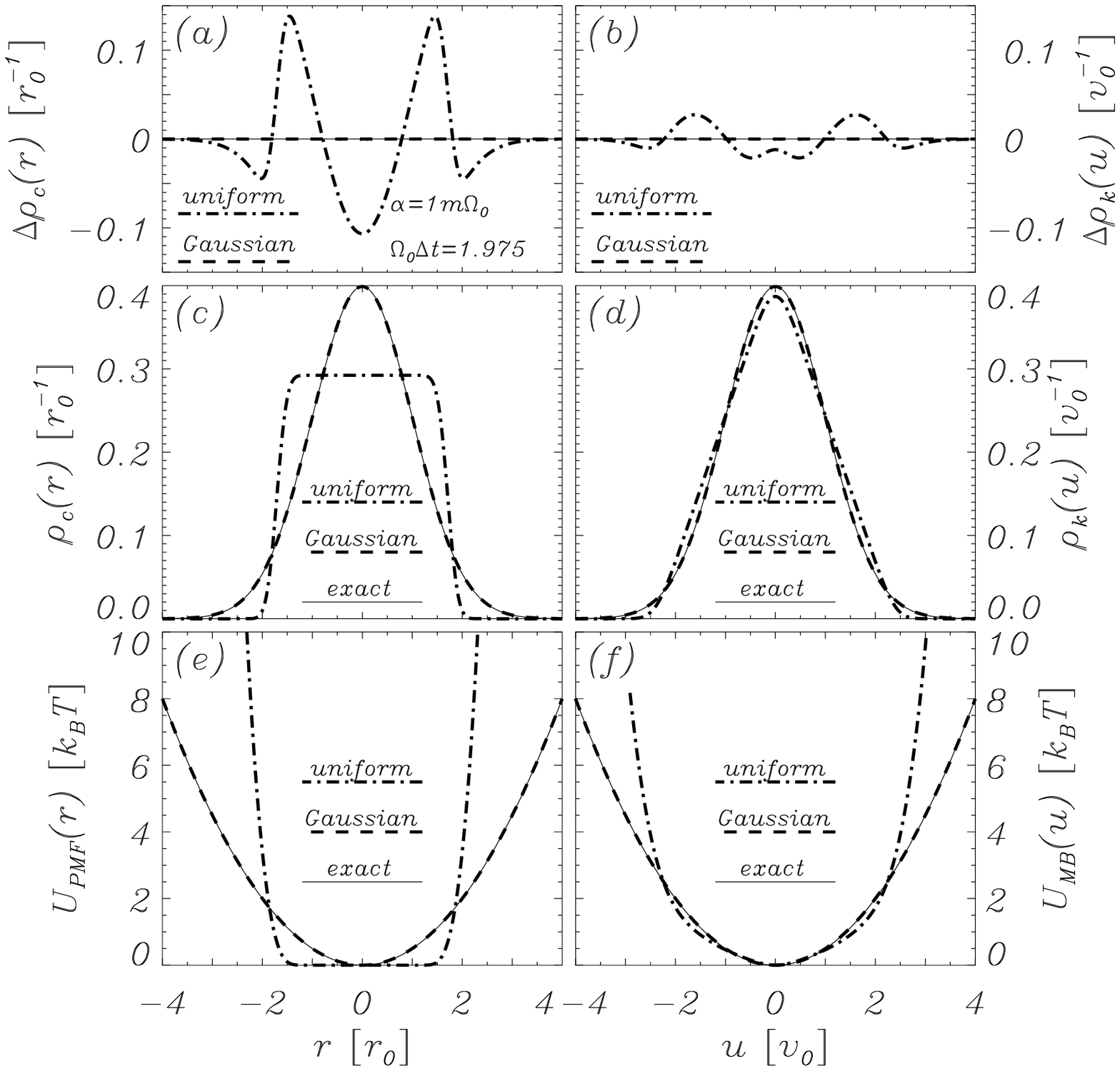}}
\caption{Configurational and kinetic distributions $\rho_c(r^n)$ and $\rho_k(u^{n+\frac{1}{2}})$ for simulations with Gaussian and uniform noise applied to the linear system Eqs.~(\ref{eq:GJ_combo}) and (\ref{eq:f_Hooke}) with $\alpha=1m\Omega_0$, $\Omega_0\Delta{t}=1.975$, $k_BT=E_0$. Shown curves are Eq.~(\ref{eq:Boltz}) (solid, labeled exact), simulations data using Gaussian noise $\beta_G^n$ (dashed, labeled Gaussian), and simulation data using uniform noise $\beta_U^n$ (dash-dot, labeled uniform). Distributions from simulations using Gaussian noise are indistinguishable from the exact references Eq.~(\ref{eq:Boltz}). (c) and (d) show the distributions, (a) and (b) show the deviations from the exact distributions Eq.~(\ref{eq:Boltz}), and (e) and (f) show the effective potential (PMF) and kinetic (MB) potentials (Eq.~(\ref{eq:PMF_MB})) derived from the distributions.
}
\label{fig:fig_5}
\end{figure}

\begin{figure}[t]
\centering
\scalebox{0.45}{\centering \includegraphics[trim={1.5cm 2.0cm 1cm 1.5cm},clip]{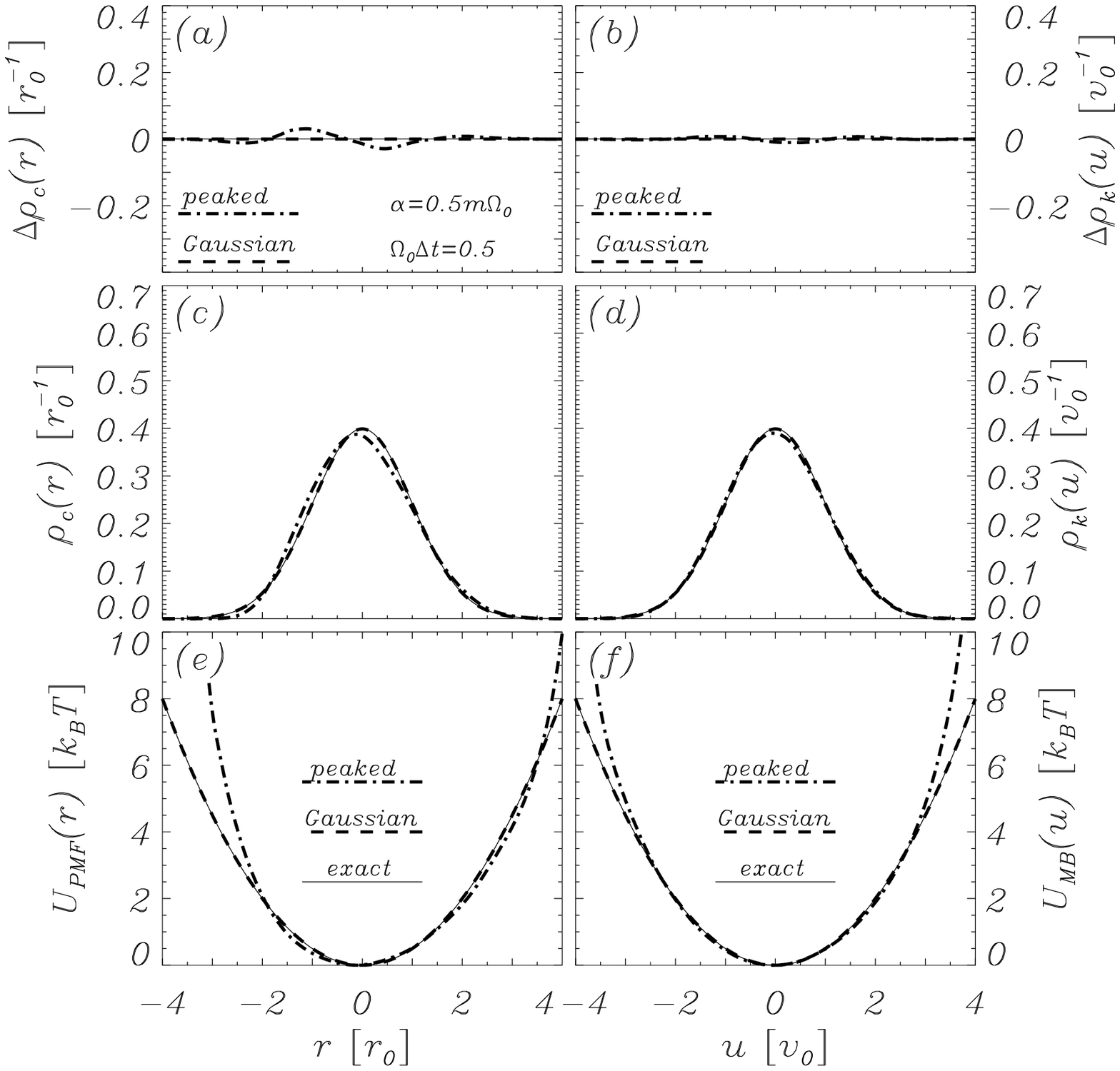}}
\caption{Configurational and kinetic distributions $\rho_c(r^n)$ and $\rho_k(u^{n+\frac{1}{2}})$ for simulations with Gaussian and peaked noise applied to the linear system Eqs.~(\ref{eq:GJ_combo}) and (\ref{eq:f_Hooke}) with $\alpha=0.5m\Omega_0$, $\Omega_0\Delta{t}=0.5$, $k_BT=E_0$. Shown curves are Eq.~(\ref{eq:Boltz}) (solid, labeled exact), simulations data using Gaussian noise $\beta_G^n$ (dashed, labeled Gaussian), and simulation data using peaked noise $\beta_P^n$ (dash-dot, labeled uniform). Distributions from simulations using Gaussian noise are indistinguishable from the exact references Eq.~(\ref{eq:Boltz}). (c) and (d) show the distributions, (a) and (b) show the deviations from the exact distributions Eq.~(\ref{eq:Boltz}), and (e) and (f) show the effective potential (PMF) and kinetic (MB) potentials (Eq.~(\ref{eq:PMF_MB})) derived from the distributions.
}
\label{fig:fig_7}
\end{figure}

\begin{figure}[t]
\centering
\scalebox{0.45}{\centering \includegraphics[trim={1.5cm 2.0cm 1cm 1.5cm},clip]{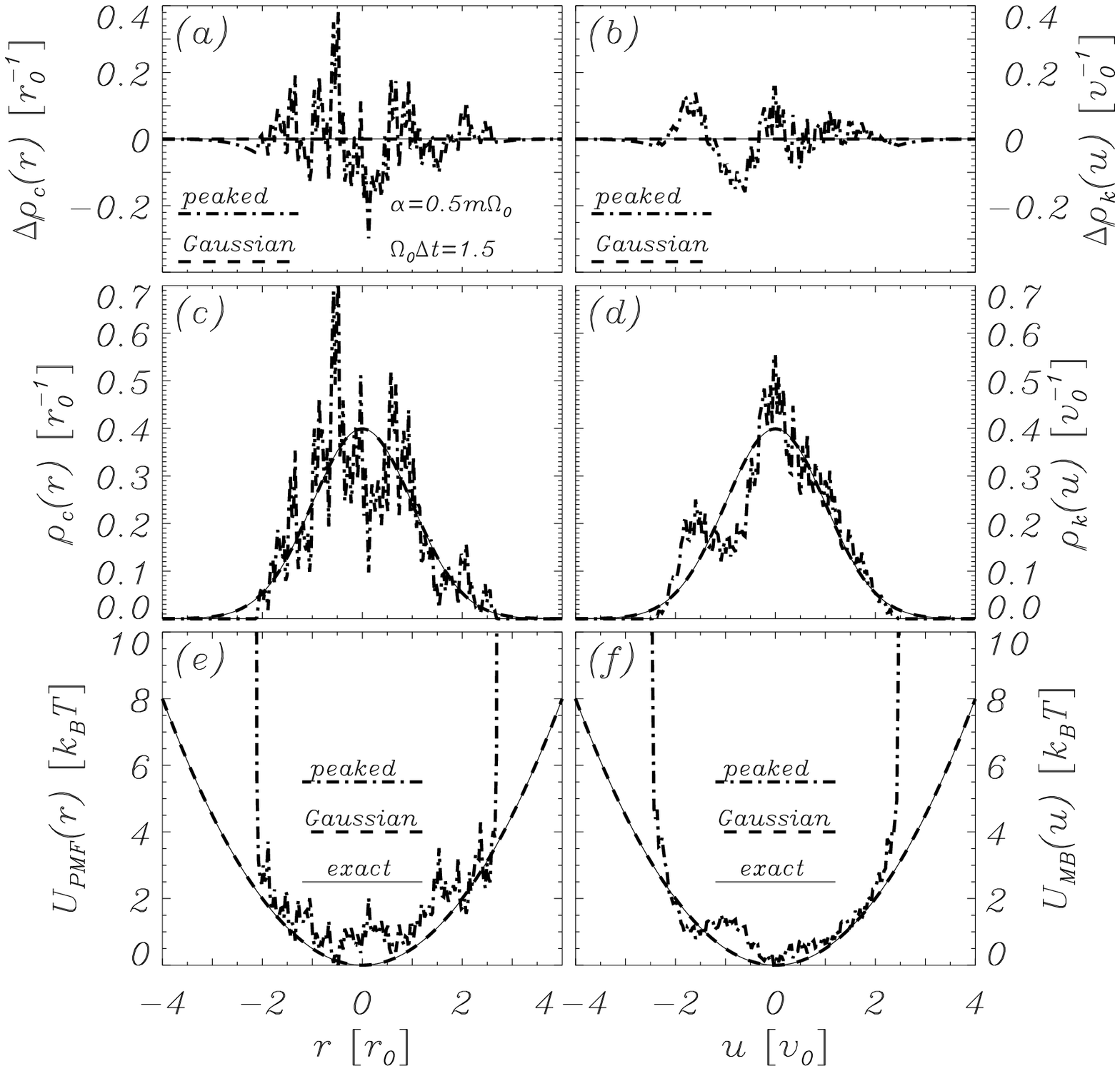}}
\caption{Configurational and kinetic distributions $\rho_c(r^n)$ and $\rho_k(u^{n+\frac{1}{2}})$ for simulations with Gaussian and uniform noise applied to the linear system Eqs.~(\ref{eq:GJ_combo}) and (\ref{eq:f_Hooke}) with $\alpha=0.5m\Omega_0$, $\Omega_0\Delta{t}=1.5$, $k_BT=E_0$. Shown curves are Eq.~(\ref{eq:Boltz}) (solid, labeled exact), simulations data using Gaussian noise $\beta_G^n$ (dashed, labeled Gaussian), and simulation data using peaked noise $\beta_P^n$ (dash-dot, labeled uniform). Distributions from simulations using Gaussian noise are indistinguishable from the exact references Eq.~(\ref{eq:Boltz}). (c) and (d) show the distributions, (a) and (b) show the deviations from the exact distributions Eq.~(\ref{eq:Boltz}), and (e) and (f) show the effective potential (PMF) and kinetic (MB) potentials (Eq.~(\ref{eq:PMF_MB})) derived from the distributions.
}
\label{fig:fig_8}
\end{figure}

\begin{figure}[t]
\centering
\scalebox{0.45}{\centering \includegraphics[trim={1.5cm 2.0cm 1cm 1.5cm},clip]{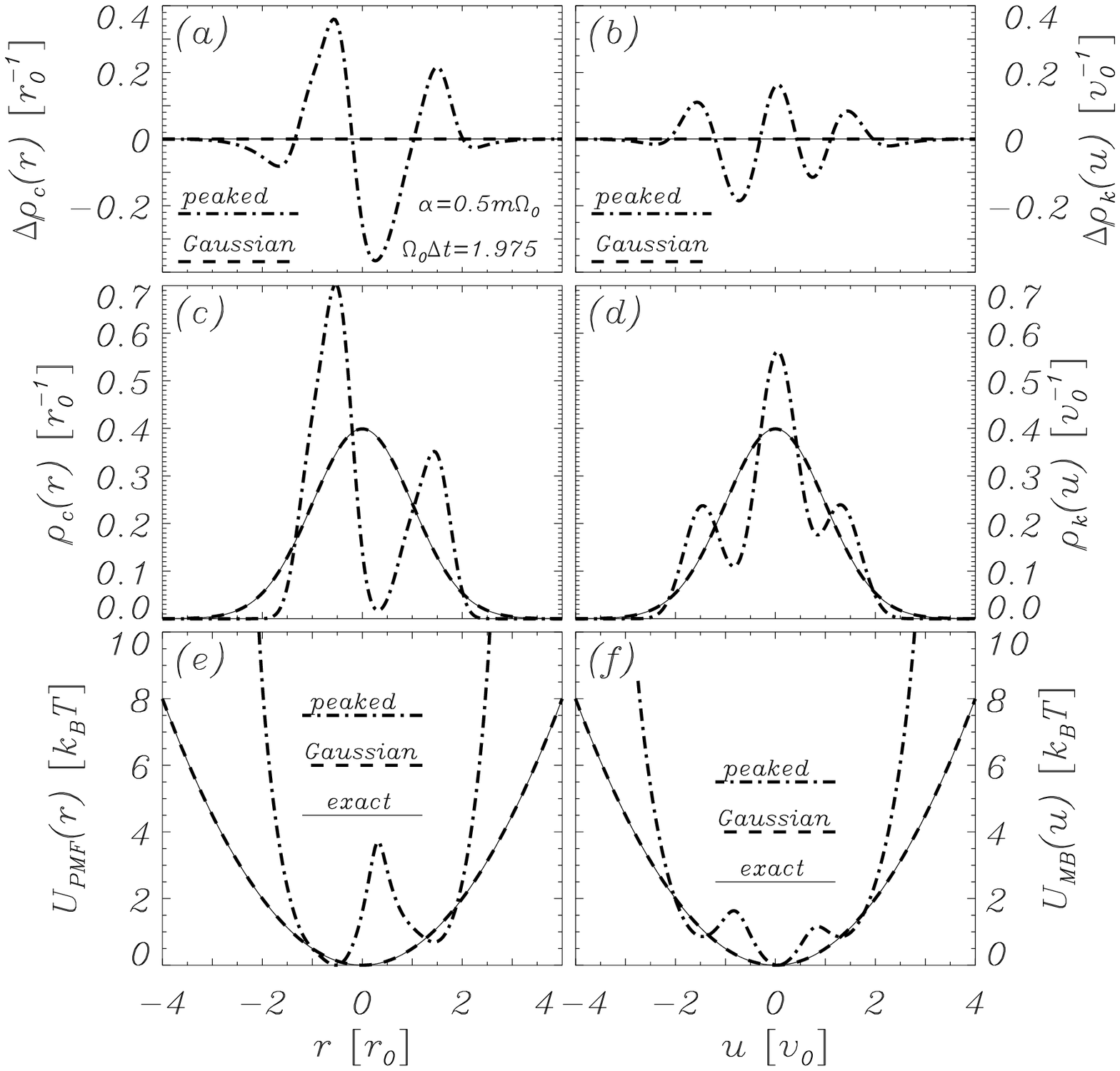}}
\caption{Configurational and kinetic distributions $\rho_c(r^n)$ and $\rho_k(u^{n+\frac{1}{2}})$ for simulations with Gaussian and uniform noise applied to the linear system Eqs.~(\ref{eq:GJ_combo}) and (\ref{eq:f_Hooke}) with $\alpha=0.5m\Omega_0$, $\Omega_0\Delta{t}=1.975$, $k_BT=E_0$. Shown curves are Eq.~(\ref{eq:Boltz}) (solid, labeled exact), simulations data using Gaussian noise $\beta_G^n$ (dashed, labeled Gaussian), and simulation data using peaked noise $\beta_P^n$ (dash-dot, labeled uniform). Distributions from simulations using Gaussian noise are indistinguishable from the exact references Eq.~(\ref{eq:Boltz}). (c) and (d) show the distributions, (a) and (b) show the deviations from the exact distributions Eq.~(\ref{eq:Boltz}), and (e) and (f) show the effective potential (PMF) and kinetic (MB) potentials (Eq.~(\ref{eq:PMF_MB})) derived from the distributions.
}
\label{fig:fig_9}
\end{figure}

\section{Simulations}
\label{sec:SimDisc}
While the above expressions provide both direct and indirect means for evaluating the thermodynamically important first four moments of the configurational and kinetic variables, we test these results against numerical simulations such that both simulations and analysis can be mutually verified. Further, the simulations provide additional details on the actual distributions $\rho_c(r^n)$ and $\rho_k(u^{n+\frac{1}{2}})$, which in some cases reveal significant deviations from Boltzmann statistics when the fluctuations $\beta^n$ are non-Gaussian.

We show results of three different distributions for the fluctuations $\beta^n$, all satisfying the discrete-time fluctuation-dissipation balance that constrains the first two moments given in Eq.~(\ref{eq:FD_d}), and thus, all resulting in correct first and second moments of $r^n$ and $u^{n+\frac{1}{2}}$, leading to correct temperature as measured by the kinetic and potential energies (see Sec.~\ref{sec:M1M2}):
\begin{description}
\begin{subequations}
\item[{\underline{Gaussian}}] $\beta^n=\beta_G^n$ with distribution $\rho_G(\beta)$:
\begin{eqnarray}
\rho_G(\beta) & = & \frac{1}{\sqrt{2\pi\langle\beta^n\beta^n\rangle}}\,\exp\left(-\frac{\beta\beta}{2\langle\beta^n\beta^n\rangle}\right) \, , \label{eq:Gauss}
\end{eqnarray}
with skewness and kurtosis
\begin{eqnarray}
\Gamma_3 & = & 0\\
\Gamma_4 & = & 3\, . 
\end{eqnarray}\label{eq:Dist_Gauss}
\end{subequations}
\begin{subequations}
\item[{\underline{Uniform}}] $\beta^n=\beta_U^n$ with distribution $\rho_U(\beta)$:
\begin{eqnarray}
\rho_U(\beta) & = & \frac{1}{2\sqrt{3\langle\beta^n\beta^n\rangle}}\times\left\{\begin{array}{ccc}
1 & , & -\sqrt{3}\,\le\,\frac{\beta}{\sqrt{\langle\beta^n\beta^n\rangle}}\,<\,\sqrt{3}\\
0 & , & {\rm otherwise}
\end{array}\right. , \nonumber \\ \label{eq:Flatn}
\end{eqnarray}
with skewness and kurtosis
\begin{eqnarray}
\Gamma_3 & = & 0\\
\Gamma_4 & = & \frac{9}{5} \, . 
\end{eqnarray}\label{eq:Dist_Flatn}
\end{subequations}
\begin{subequations}
\item[{\underline{Peaked}}] $\beta^n=\beta_P^n$ with distribution $\rho_P(\beta)$:
\begin{eqnarray}
\rho_P(\beta) & = & \frac{2}{3}\,\delta\left(\frac{\beta}{\sqrt{\langle\beta^n\beta^n\rangle}} + \frac{1}{\sqrt{2}}\right)\nonumber \\
& + & \frac{1}{3}\,\delta\left(\frac{\beta}{\sqrt{\langle\beta^n\beta^n\rangle}}-\sqrt{2}\right)\, , \label{eq:Peakn}
\end{eqnarray}
with skewness and kurtosis
\begin{eqnarray}
\Gamma_3 & = & \frac{1}{\sqrt{2}}\\
\Gamma_4 & = & \frac{12}{5}\, .
\end{eqnarray}\label{eq:Dist_Peakn}
\end{subequations}
\end{description}
The three distributions are sketched in Fig.~\ref{fig:fig_1}, and the realizations of the pseudo-random numbers $\beta^n$ are produced from the uniformly distributed numbers in $[0,1)$ generated by the RANMAR algorithm (used in LAMMPS \cite{LAMMPS-Manual}).

In order to validate the analysis of the previous section, we simulated the noisy harmonic oscillator Eqs.~(\ref{eq:GJ_combo}) and (\ref{eq:f_Hooke}) with several of the GJ methods, which all give the same statistical results for Gaussian noise once the time step is appropriately scaled according to the method. For simplicity we limit the displayed results to those of the GJ-I (GJF-2GJ) method \cite{2GJ,GJ}, for which
\begin{subequations}
\begin{eqnarray}
c_2 & = & \frac{1-\frac{\alpha\Delta{t}}{2m}}{1+\frac{\alpha\Delta{t}}{2m}} \\
c_1 \; = \; c_3 & = & \frac{1}{1+\frac{\alpha\Delta{t}}{2m}} \, .
\end{eqnarray}
\end{subequations}
The GJF-2GJ method overlaps in the configurational coordinate with the GJF method \cite{GJF1}. We note that the GJ-I (GJF-2GJ) method is available for use in LAMMPS \cite{LAMMPS-Manual}. With time normalized by the characteristic unit $t_0$ given by the inverse of the natural frequency $\Omega_0$, the reduced damping is $\alpha/m\Omega_0$, and the characteristic energy is $E_0=\kappa r_0^2$, where $r_0$ is a chosen characteristic length to which $r$ is normalized. The simulations are conducted for $k_BT=E_0$ for a variety of values of $\alpha/m\Omega_0$ in the entire stability range $\Omega_0\Delta{t}<2$. 
Each statistical value is calculated from averages over 1,000 independent simulations, each with 10$^8$ time steps.

Figure~\ref{fig:fig_2} shows the results of using the uniform noise distribution $\beta_U^n$ (Eq.~(\ref{eq:Flatn})) alongside the results of using the Gaussian $\beta_G^n$ (Eq.~(\ref{eq:Gauss})) as reference. Figures~\ref{fig:fig_2}ab, respectively displaying results for configurational and kinetic variables, verify the general result from Sec.~\ref{sec:M1M2}; namely that a statistical property, which depends only on the first and second moments of the noise distribution, will be correctly evaluated for the GJ methods if the noise satisfies Eq.~(\ref{eq:FD_d}). The shown data on each plot are for damping parameters $\alpha/m\Omega_0$=0.1, 0.25, 0.5, 1, 2.5, 5, and the results for both Gaussian and uniform noise distributions are clearly in close agreement with the expected values given in Eqs.~(\ref{eq:Ep_ave}) and (\ref{eq:Ek_ave}). This has previously been extensively validated for Gaussian noise \cite{2GJ,GJ}.

Figure~\ref{fig:fig_2}cd displays simulation results and comparisons to the expectations from the analysis in Sec.~\ref{sec:M4} for the configurational and kinetic energy fluctuations $\sigma_p$ and $\sigma_k$ found in Eqs.~(\ref{eq:sigma_p}) and (\ref{eq:sigma_k}). Since the Gaussian noise $\beta_G^n$ from Eq.~(\ref{eq:Gauss}) applied to the linear system produces Gaussian distributions for $r^n$ and $u^{n+\frac{1}{2}}$ with correct, and time-step independent, first and second moments (see Sec.~\ref{sec:M1M2}), it follows that the fourth moments are also correct and time-step independent. This is clearly observed on the energy fluctuation figures, where both simulation results and the results of solving Eqs.~(\ref{eq:Ar4R})-(\ref{eq:R}) for $\Gamma_4=3$ are shown to be indistinguishable and constant at the correct values $\sigma_p=\sigma_k=k_BT/\sqrt{2}$. However, the uniformly distributed noise $\beta_U^n$ (Eq.~(\ref{eq:Flatn})) produces neither Gaussian nor uniform distributions for $r^n$ and $u^{n+\frac{1}{2}}$, but instead produces distributions that depend on the time step and the damping parameter, even if Eq.~(\ref{eq:FD_d}) is satisfied. This can be seen in the figure for the energy fluctuations, Fig.~\ref{fig:fig_2}cd, where both simulation results and the results of Eqs.~(\ref{eq:Ar4R})-(\ref{eq:R}) for $\Gamma_4=3$ and $\Gamma_4=\frac{9}{5}$, the latter being the kurtosis for the uniform distribution, are shown. The simulation results are indistinguishable from the analysis. It is obvious that when using the uniform noise, the energy fluctuations, which depend on the fourth moments of the variables, can deviate significantly from the correct value, given by the Gaussian noise. We also observe that the fluctuations approach the correct values for small time steps or small damping parameters. This is understandable since the damping per time step in those limits is very small, and the fluctuations in $r^n$ and $u^{n+\frac{1}{2}}$ therefore are composed of $\beta_U^n$ contributions from many time steps, allowing the integrated noise in $r^n$ and $u^{n+\frac{1}{2}}$ to become near-Gaussian by the central limit theorem. Conversely, if the damping per time step becomes appreciable, then the effective noise in $r^n$ and $u^{n+\frac{1}{2}}$ will be composed by only a few $\beta_U^n$ contributions, which do not approximate a Gaussian outcome very well.

In order to look at the details of the simulated distributions, we have selected a few representative parameter values as illustrations. Figures~\ref{fig:fig_3}-\ref{fig:fig_5} show the simulated configurational and kinetic distributions, $\rho_c(r)$ and $\rho_k(u)$, for $\alpha=1m\Omega_0$ and $\Omega_0\Delta{t}=0.5, 1.5, 1.975$, respectively, for simulations using Gaussian (dashed) and uniform (dash-dotted) noise distributions. Also shown are the results from the exact Gaussian distributions, $\rho_{k,e}(u)$ and $\rho_{c,e}(r)$, (solid, from Eqs.~(\ref{eq:MBoltzmann_exact}) and (\ref{eq:Boltzmann_c_exact})) that are expected from continuous-time Langevin dynamics. The distributions, $\rho_c(r^n)$ and $\rho_k(u^{n+\frac{1}{2}})$, are shown in Figures~\ref{fig:fig_3}-\ref{fig:fig_5}cd, the deviations,
\begin{subequations}
\begin{eqnarray}
\Delta\rho_c(r^n) & = & \rho_c(r^n)-\rho_{c,e}(r^n)\\
\Delta\rho_k(u^{n+\frac{1}{2}}) & = & \rho_k(u^{n+\frac{1}{2}})-\rho_{k,e}(u^{n+\frac{1}{2}}) \, ,
\end{eqnarray}
\end{subequations}
 from the expected (correct) distributions are shown in Figures~\ref{fig:fig_3}-\ref{fig:fig_5}ab, and the effective configurational and kinetic potentials,
 \begin{subequations}
 \begin{eqnarray}
U_{PMF}(r^n) & = & -k_BT\ln\rho_c(r^n)+\tilde{{\cal C}}_c\\
U_{MB}(u^{n+\frac{1}{2}}) & = & -k_BT\ln\rho_k(u^{n+\frac{1}{2}})+\tilde{{\cal C}}_k\, ,
 \end{eqnarray}\label{eq:PMF_MB}\noindent
 \end{subequations}
where $\tilde{{\cal C}}_c$ and $\tilde{{\cal C}}_k$ are determined such that ${\rm min}[U_{PMF}(r^n)]={\rm min}[U_{MB}(u^{n+\frac{1}{2}})]=0$, are shown in Figures~\ref{fig:fig_3}-\ref{fig:fig_5}ef. For the harmonic potential, the statistically correct values of these potentials should be $\frac{1}{2}\kappa(r)^2$ and $\frac{1}{2}m(u)^2$, which are indicated by thin solid curves. For $\Omega_0\Delta{t}=0.5$, Figure~\ref{fig:fig_3} shows that even a seemingly modest deviation from a pure Gaussian distribution can have rather large impacts on fourth-moment thermodynamic measures, which for these parameters in Fig.~\ref{fig:fig_2}cd are seen to result in configurational and kinetic energy fluctuations being depressed by about 7\% and 14\%, respectively. Increasing the time step to $\Omega_0\Delta{t}=1.5$ amplifies the deformation of the $\beta_U^n$-generated distributions away from the Gaussian. As seen in Fig.~\ref{fig:fig_4}ef, the increasingly noticeable difference is that the uniform noise yields a more confined exploration of phase-space than the Gaussian distribution does, consistent with the depression in the fourth moment seen in Fig.~\ref{fig:fig_2}cd. Finally, in Fig.~\ref{fig:fig_5}, we show the results for a time step, $\Omega_0\Delta{t}=1.975$, very close to the stability limit. This extreme case shows that the uniform distribution in noise also can produce a near uniform distribution for $r^n$, while the distribution for $u^{n+\frac{1}{2}}$ is near triangular, consistent with a sum of two uniformly distributed numbers contributing to the kinetic fluctuations. Again, we see from Fig.~\ref{fig:fig_5}ef that the sampling of the phase space is much more limited when using uniform noise than when using Gaussian, even if the measured temperatures, configurational as well as kinetic, are measured to the correct values.

Following the spirit of Refs.~\cite{Greiner_1,Dunweg_1} we explore the application of a more challenging noise distribution $\beta_P^n$ (the peaked distribution defined in Eq.~(\ref{eq:Peakn})), which is both discrete and asymmetric (see Fig.~\ref{fig:fig_1}). As was the case for the uniform noise discussed above, we also here reference the results next to those of Gaussian noise, which give time-step independent and correct behavior. Conducting numerical simulations of Eqs.~(\ref{eq:GJ_combo}) and (\ref{eq:f_Hooke}) for different time steps and for normalized friction $\alpha/m\Omega_0=0.1, 0.25, 0.5, 1, 2.5, 5$ as described above, and comparing to the evaluation of the moments from the analyses of Sec.~\ref{sec:moments}, we obtain the data shown in Fig.~\ref{fig:fig_6}, where the simulation results are indistinguishable from the results of the analyses for third and fourth moments in Secs.~\ref{sec:M3} and \ref{sec:M4}. As expected from the analysis, the results of the second moments shown in Fig.~\ref{fig:fig_6}ab of $r^n$ and $u^{n+\frac{1}{2}}$, namely the potential and kinetic energies, are perfectly aligned with statistical mechanics, since the peaked distribution for $\beta_P^n$ satisfies the two moments in Eq.~(\ref{eq:FD_d}). Since the applied noise is here asymmetric, the resulting third moments of $r^n$ and $u^{n+\frac{1}{2}}$ may also become non-zero for non-zero time steps. This is seen in Figures~\ref{fig:fig_6}cd, where we observe rather complex behavior as a function of the system parameters. We do, however, see that for $\Omega_0\Delta{t}\rightarrow0$ the third moments approach zero, consistent with the central limit theorem that guarantees a Gaussian outcome when a very large number of noise values contribute to the variables. Finally, in comparison with Fig.~\ref{fig:fig_2}cd for uniform noise, we see very similar behavior for the energy fluctuations (fourth moments) for the application of the peaked distribution in Fig.~\ref{fig:fig_6}ef. Even if the uniform and peaked distributions are very different in appearance, the similarities between their outcomes in their fourth moments are not surprising, given that the analysis for the fourth moments in Sec.~\ref{sec:M4} shows that the difference between the two only depends on the kurtoses, which have the values $\Gamma_4=\frac{9}{5}$ (uniform) and $\Gamma_4=\frac{12}{5}$ (peaked). Again, we find that all signatures of non-Gaussian noise vanish for $\Omega_0\Delta{t}\rightarrow0$, and we find that non-zero time steps result in depressions of the thermodynamically important energy fluctuations, as derived from the fourth moments of the configurational and kinetic coordinates.

The details of the simulated coordinate distributions arising from the peaked noise are exemplified in Figs.~\ref{fig:fig_7}-\ref{fig:fig_9} for $\alpha/m\Omega_0=0.5$ and $\Omega_0\Delta{t}=0.5, 1.5, 1.975$, respectively. For the smaller of the time steps, shown in Fig.~\ref{fig:fig_7}, we see the seemingly modest skewness and deformations of the coordinate distributions arising from the peaked noise. Yet, these modest deformations are what provide the somewhat significant deviations from Gaussian characteristics in third and fourth moments seen in Fig.~\ref{fig:fig_6}cd and Fig.~\ref{fig:fig_6}ef for $\Omega_0\Delta{t}=0.5$. More dramatic deviations from Gaussian/Boltzmann characteristics are found in a large range of $\Omega_0\Delta{t}$, including the value $\Omega_0\Delta{t}=1.5$ shown in Fig.~\ref{fig:fig_8}. The seemingly discontinuous distribution is not a result of insufficient statistics. Rather, it is the interference between the discrete nature of the applied noise distribution with the discrete time step that happens to distinctively select certain preferred values of $r^n$ and $u^{n+\frac{1}{2}}$ over others. It is noticeable that these types of peculiar distributions appear without any obvious  or abrupt signatures in the first four moments, as seen in Fig.~\ref{fig:fig_6}. As the time step $\Omega_0\Delta{t}=1.975$ is pushed close to the stability limit, we again find smooth coordinate distributions, seen in Fig.~\ref{fig:fig_9}, where the applied peaked noise is visible throughout the different displays. In analogy with the visualization of the distributions from the uniform noise, this is the limit where the friction per time step is relatively large, thereby making the behavior of the coordinates $r^n$ and $u^{n+\frac{1}{2}}$ subject to only a few noise contributions at a time.

\section{Discussion}
\label{sec:discussion}
In light of recent advances in stochastic thermostats for simulating Langevin equations with accurate statistics across the stability range of the applied time step when using Gaussian noise \cite{2GJ,GJ}, we have revisited the investigations of advantages and disadvantages of using non-Gaussian thermal noise in discrete time. Given the systemic first and second order time-step errors of the traditional methods (e.g., Refs.~\cite{BBK,SS}), which necessitated rather small time steps for accurate simulations, it was previously concluded \cite{Greiner_1,Dunweg_1} that other distributions, including the desirable uniform distribution, would be efficient substitutions since the central limit theorem would ensure near-Gaussian outcomes for small time steps, thereby not further significantly distort the simulation results due to the imperfect noise. This result has been a very useful tool over the years when computational efficiency could benefit from not converting stochastic variables from uniform to Gaussian. However, as we have demonstrated in this paper, when the time step becomes large, which is allowed by the modern GJ methods when using Gaussian noise, the application of non-Gaussian noise does not retain the time-step independent benefits of these methods in thermodynamic measures that involve moments higher than the second. When the applied noise conforms to Eq.~(\ref{eq:FD_d}), we have found that the GJ methods are invariant to the specific noise distributions in measures of first and second moments, such as configurational and kinetic temperatures. Thus, it is deceiving to judge the quality of the thermostat based on those moments alone. As is evident from the third and fourth moments, as well as the visual impressions of the actual distributions of $r^n$ and $u^{n+\frac{1}{2}}$, the sampling of the phase-space can be quite distorted (non-Boltzmann) even if the measured temperatures yield correct values. All of these results are consistent with the significance of how the noise must be defined in discrete-time; namely through the integral Eq.~(\ref{eq:beta_n}), which ensures that any underlying distribution for $\beta(t)$ will yield a Gaussian outcome for $\beta^n$ due to the central limit theorem.  It follows that applying any other distribution than Gaussian in discrete time is formally invalid, and certainly leads to significant sampling errors unless $\alpha\Delta{t}/m$ is small enough for the one-time-step velocity attenuation parameter to be $c_2\approx1$. It is therefore the conclusion of this work that Gaussian noise must be applied to the modern stochastic integrators if one wishes to take advantage of the large time-step benefits of their properties.

\section{Acknowledgments}
The author is grateful to Charlie Sievers for sharing LAMMPS simulation results using the GJF-2GJ method, supporting the conclusions of this paper for more complex molecular dynamics systems, and to Lorenzo Mambretti for assistance with the RANMAR random number generator. The author is also grateful for initial discussions with Chungho Cheng on non-Gaussian noise.

\section{Data Availability Statement}
The data presented and discussed in the current study are available from the author on reasonable request.

\section{Funding and/or Competing Interests}
No funding was received for conducting this study.
The author has no relevant financial or non-financial interests to disclose.



\begin{thebibliography}{99}
\bibitem{comment_symmetry} Although not necessary, it is generally reasonable to expect that the distribution of $\beta(t)$ is symmetric, yielding all odd moments zero.
\end{thebibliography}
\end{document}